\documentclass[prd,superscriptaddress,unsortedaddress,twocolumn,%
  showpacs,preprintnumbers,amsmath,amssymb,]{revtex4-1}
\usepackage{graphicx}
\usepackage[breaklinks,colorlinks=true]{hyperref}

\newcommand{\calO}{\mathcal{O}}
\newcommand{\be}{\boldsymbol{e}}
\newcommand{\bj}{\boldsymbol{j}}
\newcommand{\bp}{\boldsymbol{p}}
\newcommand{\bv}{\boldsymbol{v}}
\newcommand{\bx}{\boldsymbol{x}}
\newcommand{\bB}{\boldsymbol{B}}
\newcommand{\bE}{\boldsymbol{E}}
\newcommand{\bbeta}{\boldsymbol{\beta}}
\newcommand{\bnabla}{\boldsymbol{\nabla}}
\newcommand{\bOmega}{\boldsymbol{\Omega}}

\begin{document}
\title{Boost invariant formulation of the chiral kinetic theory}
\author{Shu Ebihara}
\author{Kenji Fukushima}
\author{Shi Pu}
\affiliation{Department of Physics, The University of Tokyo, %
             7-3-1 Hongo, Bunkyo-ku, Tokyo 113-0033, Japan}
\begin{abstract}
  We formulate the chiral kinetic equation with the longitudinal boost
  invariance.  We particularly focus on the physical interpretation of
  the particle number conservation.  There appear two terms associated
  with the expansion, which did not exist in the non-chiral kinetic
  equation.  One is a contribution to the transverse current arising
  from the side-jump effect, and the other is a change in the density
  whose flow makes the longitudinal current.  We point out a
  characteristic pattern in the transverse current driven by the
  expansion, which we call the chiral circular displacement.
\end{abstract}
\maketitle

\section{Introduction}

In theoretical and experimental research on relativistic heavy-ion
collisions and also condensed matter systems of Weyl/Dirac semimetals
chiral matter with massless fermionic dispersions has been attracting
more and more interest.  The most notable features in chiral matter
are exotic phenomena driven by chirality and quantum anomaly such as
the Chiral Magnetic Effect
(CME)~\cite{Kharzeev:2007jp,Fukushima:2008xe};  the CME is realized as
an electric current along the external magnetic field in the presence
of imbalanced chiral charge (see recent
reviews~\cite{Kharzeev:2015znc,Hattori:2016emy} for more details).
The CME physics
has triggered various theoretical
investigations including the anomalous hydrodynamics with chiral
anomaly~\cite{Son:2009tf,Sadofyev:2010pr,Pu:2010as}, lattice numerical
simulations of quantum chromodynamics (QCD) with a chiral chemical
potential~\cite{Abramczyk:2009gb,Buividovich:2009wi,Buividovich:2009zzb,%
Buividovich:2010tn,Yamamoto:2011gk},
holographic models based on gauge/gravity
correspondence~\cite{Erdmenger2009,Torabian2009a,Banerjee2011}, the
quantum kinetic description with Wigner
functions~\cite{Gao:2012ix,Gao:2017rgi}, and the kinetic theory with
the Berry
curvature~\cite{Son:2012wh,Son:2012zy,Stephanov:2012jts,Chen:2012ca}.

In non-central heavy-ion collisions, a magnetic field as strong as the
QCD energy scales is expected;  see e.g.\
Ref.~\cite{Skokov:2009qp} for the first estimate using the
Ultrarelativistic Quantum Molecular Dynamics (UrQMD) model,
Refs.~\cite{Bzdak:2011yy,Deng:2012pc,Roy:2015coa} for even-by-event
simulations, and Refs.~\cite{Tuchin:2014iua,Li:2016tel} for the case
with finite chiral conductivity.  Moreover, with imbalanced chirality
induced from the topological transition on P- and CP-violating gauge
backgrounds, the CME currents are anticipated to generate the charge
asymmetry fluctuations as measured first in the
STAR Collaboration~\cite{Abelev:2009ac,Abelev:2009ad}.
To distinguish the genuine signal from flow backgrounds,
quantitative studies are indispensable.  Besides, some other novel
effects from chiral transport may also play a substantial role in
quantitative simulations for the heavy-ion collisions, e.g.\ the
chiral electric separation effect~\cite{Huang:2013iia,Pu:2014cwa}, the
chiral Hall separation effect~\cite{Gursoy:2014aka,Pu:2014fva}, and
higher order effects~\cite{Chen:2016xtg, Gorbar:2016qfh}.

For real-time simulations of the CME, three approaches are commonly
adopted.  The first one is the classical statistical simulation based
on the solution of the classical Yang-Mills and Dirac equations;  see
Refs.~\cite{Mace:2016svc, Mace:2016shq, Berges:2017igc} for recent
studies.  The second is the anomalous hydrodynamics or the chiral
magneto-hydrodynamics (CMHD), see
Ref.~\cite{Hirono:2014oda,Jiang:2016wve} for recent studies and also
Refs.~\cite{Roy:2015kma,Pu:2016ayh} for the analytic one-dimensional
solutions, Ref.~\cite{Pu:2016bxy} for the two-dimensional Bjorken
expanding case, and Ref.~\cite{Inghirami:2016iru} for more
simulations.

When the occupation number is dilute, the third and the
most microscopic approach becomes feasible, i.e.\ the kinetic theory
that can be applied for general systems out of equilibrium.  The
conventional Boltzmann equation is not capable of treating the spin
dynamics~\cite{Pu:2010as} and the kinetic theory augmented with the
spin dynamics for chiral matter is called the chiral kinetic theory
(CKT)~\cite{Son:2012wh,Son:2012zy,Stephanov:2012jts,Chen:2012ca}.
See also Refs.~\cite{Sun:2016nig,Huang:2017tsq} for latest CKT
simulations.  To derive the equations of motion for the spin, the
Berry curvature~\cite{Berry1984} should be added to the action in the
path-integral formulation (see Refs.~\cite{Xiao:2005eif,Duval2006} for
discussions and Ref.~\cite{Xiao2010} for a review).  In this way the
Berry curvature corrections are expected to the Boltzmann equation, as
was concluded in
Refs.~\cite{Son:2012wh,Son:2012zy,Stephanov:2012jts,Chen:2012ca,%
  Chen:2013iga,Chen:2014cla,Hidaka:2016yjf}.

To consider the early stage of the time evolution in the relativistic
heavy-ion collision, the Boltzmann equation in expanding geometry has
played an essential role.  One of the earliest applications is found
in the explanation of the isotropization mechanism in the relaxation
time approximation~\cite{Baym:1984guh}.  The QCD interaction was taken
into account in Ref.~\cite{Mueller:1999pi}, which is the foundation
for the bottom-up thermalization scenario~\cite{Baier:2000sb} (see
Ref.~\cite{Fukushima:2016xgg} for a recent review and references
therein).  So far, most of preceding works have been focused on the
gluonic sector called ``glasma'' (see
Refs.~\cite{Fukushima:2016xgg,Blaizot:2016qgz} for pedagogical reviews
on the gluon saturation and the glasma picture) because the early time
dynamics is dominated by gluonic degrees of freedom due to quantum
evolution.  Here, let us point out  advantages to consider the quark
sector. 

The first important feature in the quark sector is that, unlike
gluons, physical conditions validate the kinetic description for
quarks.  The reason why the quantum Boltzmann equation~\footnote{We
  use a word ``quantum'' in a sense of discussions in
  Refs.~\cite{Epelbaum:2014mfa,Fukushima:2017odi}.}
cannot be utilized in the glasma regime is that the 
gluons are overpopulated and the non-linearity must be fully taken
into account.  In contrast, quarks are never overpopulated since they
obey the Fermi statistics.

The second is that the quark distribution in the very early stage of
the heavy-ion collision should be dilute, so that the dominant process
should be the interaction between quarks and background gluonic
fields.  Thus, even in an approximation to neglect the collision
integral, the CKT can correctly capture the quark dynamics, and
nevertheless, the results are non-trivial as revealed in this paper.

The third is that, in the non-central collision, the strong external
magnetic field quickly decays, but it can survive over the time scale
of the glasma that may accommodate the topological charge
fluctuations~\cite{Kharzeev:2001ev,Mace:2016svc}.  Such a combination
of the magnetic and the gluonic fields would yield visible effects of
the local parity violation.  For the sake of investigating this, the
CKT provides us with a powerful and microscopic tool, and in
principle, we should be able to estimate physical observables from the
glasma-type initial condition.

There is an obstacle, however.  Although the CKT itself has been
established, it is far from straightforward to apply the CKT to the
heavy-ion phenomenology since the realization of the Lorentz symmetry
in chiral matter is highly
non-trivial~\cite{Chen:2014cla,Chen:2015jop}.  Under the Lorentz
boost, usually, the coordinate vector $\bx$ and the momentum vector
$\bp$ as well as the electromagnetic fields should transform as 
vectors and tensors, but it has been found in
Refs.~\cite{Chen:2014cla,Chen:2015jop} that transformed $\bx$ and
$\bp$ must have extra terms of order of the Berry curvature, which are
called side-jump effects~\cite{Chen:2014cla}.  One can reach the same
conclusion by looking at the classical effective action of chiral
matter.  Later, from the quantum field theoretical formulation, it has
been demonstrated in Ref.~\cite{Hidaka:2016yjf} that, instead of
elaborating $\bx$ and $\bp$ transformations, a non-trivial
transformation on the distribution function under the Lorentz boost
can lead to equivalent consequences~\cite{Chen:2015jop}.

The question that we would like to address in this work is;  what is
the counterpart of the boost invariant Boltzmann equation of
Refs.~\cite{Baym:1984guh,Mueller:1999pi} for quarks?  For clarity let
us make two remarks here.  (1) One might think that the boost
invariance makes the topology of theory trivial.  This is completely
true, and the topological charge is vanishing~\cite{Kharzeev:2001ev};
however, the topological charge \textit{density} can distribute over
space, which is nothing but a microscopic picture of the local parity
violation.  (2) Also, one might think that the boost invariance is
anyway broken by quantum fluctuations.  This is indeed so in the
gluonic sector that exhibits the glasma
instability~\cite{Romatschke:2006nk} but not true in the quark sector.
As long as the time scale is shorter than the glasma instability
growth (i.e.\ up to a few times the saturation momentum inverse), we
can treat glasma fields as boost invariant gluonic backgrounds.  Also,
the spatial variations of the external magnetic field spread more
smoothly than the saturation scale, and so we can again regard the
magnetic field as approximately boost invariant.  Then, one can prove
that the mode functions in the quark sector keep boost invariance as
long as external fields have no dependence on the coordinate
rapidity~\cite{Gelis:2004jp,Gelis:2015eua}.

More specifically, spinors as solutions of the Dirac equation
explicitly have boost invariant properties once the spinor components
are appropriately transformed in the co-moving frame.  Therefore,
naturally,
the quark distribution function appearing in the CKT must share the
same properties because the distribution function is microscopically
defined in terms of the solutions of the Dirac equation.  Then, with
the boost invariant Ansatz for the distribution function, the CKT is
rewritten in a way that makes the expansion effect manifested.

The organization of the present paper is as follows;
in Sec.~\ref{sec:Dirac} we discuss the Dirac equation and the boost
invariance following the discussions in Ref.~\cite{Gelis:2015eua}.  In
Sec.~\ref{sec:BjorkenBE} we then briefly review what is known for the
ordinary Boltzmann equation with the boost invariant Ansatz for the
distribution function.  We proceed to Sec.~\ref{sec:CKT} to discuss
the application of the CKT to the system with boost invariance, which
is followed by some considerations on the non-trivial realization of
the particle number conservation in Sec.~\ref{sec:number}.  Finally, we
make a summary in Sec.~\ref{summary}.

\section{Dirac equation with boost invariance}
\label{sec:Dirac}

Let us first consider the Dirac equation under external
electromagnetic fields.  The generalization to non-Abelian fields
should be straightforward.  Throughout this work we always assume
boost invariant electromagnetic fields, which is the case in the
CGC-type chromo-fields for instance.  It is useful to rewrite the
equation of motion in terms of the Bjorken coordinates, namely, the
proper time, $\tau\equiv\sqrt{t^2-z^2}$, and the coordinate rapidity,
$\eta\equiv\frac{1}{2}\ln|(t-z)/(t+z)|$.  Then, boost invariance is
translated into the $\eta$ independence of external fields.  We can
also introduce the momentum counterparts, that is, the transverse
momentum, $p_\perp\equiv\sqrt{p_0^2-p_z^2}$, and the momentum rapidity,
$y\equiv\frac{1}{2}\ln|(p_0-p_z)/(p_0+p_z)|$.

The limit of vanishing external fields is the simplest example that is
useful for our consideration.  Needless to say, the solution of the
equation of motion is then a plane wave, $e^{ip\cdot x}$, apart from
the spinor components.  Actually, using the Bjorken coordinates, we
can express the (longitudinal part of) plane wave as
$e^{\mp i(p_0 x^0+p_z z)}=e^{\mp ip_\perp\tau\cosh(y-\eta)}$.  In this
way, we understand that the Lorentz invariance requires a special
combination of rapidity dependence through $y-\eta$ only.  Now, with
interactions introduced in the covariant derivative, the massless
Dirac equation reads~\cite{Gelis:2004jp},
\begin{equation}
  i\biggl( \gamma^0 \partial_\tau + \frac{\gamma^0}{2\tau}
    + \frac{\gamma^z}{\tau} D_\eta
    + \gamma^i D_i \biggr) \tilde{\psi} = 0\;,
\end{equation}
where $\tilde{\psi}=e^{-\frac{\eta}{2}\gamma^0\gamma^z}\psi$ and the
$A_\tau=0$ gauge is chosen.  We can even simplify the above by
changing the fermion basis as
$\tilde{\psi}\to\hat{\psi}=\tau^{1/2}\tilde{\psi}$
yielding~\cite{Gelis:2015eua},
\begin{equation}
  i\biggl( \gamma^0 \partial_\tau
    + \frac{\gamma^z}{\tau} D_\eta
    + \gamma^i D_i \biggr) \hat{\psi} = 0\;.
\end{equation}
As long as there is no $\eta$ dependence in gauge potentials, we can
now explicitly solve the $\eta$ dependence for particles and
anti-particles by plugging $e^{\mp ip_\perp\tau\cosh(y-\eta)}\chi_\mp$
for $\hat{\psi}$ into the Dirac equation, and we then find for the
right-handed two-spinor part for example;
\begin{equation}
  \biggl(\partial_\tau \mp i p_\perp e^{-\sigma^3(y-\eta)}
  + i \frac{\sigma^3}{\tau} g A_\eta + \sigma^i D_i \biggr)
  \chi_\mp^R(\tau,\bx_T;y-\eta) = 0\;.
\end{equation}
Here, the rapidity dependence appears only in the second term of the
equation of motion, and obviously, the time evolution maintains the
boost invariant property;  a complete set of wave-functions are
functions of $y-\eta$ for any $\tau$.  Therefore, naturally, the
distribution function that is to be defined by wave-functions must be
also a function of $y-\eta$.

It is crucially important to notice that, when we talk about the boost
invariance of physical states, we mean their dependence through
$y-\eta$, while we assume $\eta$ independence for external fields.

\section{Boltzmann equation with boost invariance}
\label{sec:BjorkenBE}

As a preliminary exercise for the CKT formulation with boost
invariance, let us make a brief review on the ordinary Boltzmann
equation following Ref.~\cite{Mueller:1999pi}.  The Boltzmann equation
or the Vlasov equation reads,
\begin{equation}
  \frac{\partial}{\partial t}f + \bv\cdot\frac{\partial}{\partial \bx}f
  + (\bE+\bv \times \bB)\cdot\frac{\partial}{\partial\bp}f
  = I_{{\rm coll}}[f]\;,
\end{equation}
where $f(\bp,\bx,t)$ is the distribution function in phase space, and
$\bE$, $\bB$ are electromagnetic fields.  The velocity is defined
by $\bv\equiv\partial\varepsilon_p/\partial\bp$ with
$\varepsilon_p=|\bp|$.  In what follows below we will neglect the
external fields for simplicity.
 
If we impose boost invariance along the longitudinal (i.e.\ $z$)
direction, such a condition restricts the form of the distribution
function as $f(\bp,\bx,t)=f(\bp_\perp,\bx_\perp,y-\eta,\tau)$.  This
implies that the following relations should hold;
\begin{equation}
  \frac{\partial f}{\partial t}
  =\biggl(\frac{t}{\tau}\frac{\partial}{\partial\tau}
  +\frac{z}{\tau^2}\frac{\partial}{\partial\eta}\biggr)f\;,
\end{equation}
and
\begin{equation}
  \frac{\partial f}{\partial z}
  =\biggl(-\frac{z}{\tau}\frac{\partial}{\partial\tau}
  -\frac{t}{\tau^2}\frac{\partial}{\partial\eta}\biggr)f\;.
\end{equation}
Using the velocity vector, $v_x=v_y=0$ and $v_z=p_z/p_0$, we can
readily see,
\begin{equation}
  \biggl(\frac{\partial}{\partial t}+v_z\frac{\partial}{\partial z}
  \biggr) f =
  \frac{\cosh(y-\eta)}{\cosh y} \biggl( \frac{\partial}{\partial\tau}
  -\frac{\tanh(y-\eta)}{\tau}\frac{\partial}{\partial y}\biggr) f\;,
\end{equation}
where we used $\partial f/\partial\eta = -\partial f/\partial y$.
Here, again, we see that the time evolution of the Boltzmann equation
preserves the dependence only through $y-\eta$, apart from the overall
(trivially factorizable) $y$ dependence.

It is a common strategy to consider only around the mid-rapidity
region, i.e.\ $\eta\approx 0$ and $z\approx 0$, to simplify the
equations as
\begin{equation}
  \biggl(\frac{\partial}{\partial t}+v_z\frac{\partial}{\partial z}
  \biggr) f \approx
  \biggl( \frac{\partial}{\partial t}
  -\frac{p_z}{t}\frac{\partial}{\partial p_z}\biggr) f\;.
\label{eq:mid-rapidity}
\end{equation}
We note that the above prescription does not mean that we are looking
at a single point $z=0$ but we make a systematic expansion in $z$
around $z=0$ to pick up leading terms.

Now we can easily develop intuitive understanding for the second term
in the above as follows.  Let us take the momentum integration to find
an equation for the density defined by
$n(\bx)\equiv \int_{\bp}f(\bp,\bx)$.  After the integration by part
with respect to $p_z$, we find,
\begin{equation}
  \frac{\partial n}{\partial t} + \frac{n}{t}
  + \bnabla_{\bx_\perp}\cdot \bj_\perp = 0\;,
\label{eq:continuity}
\end{equation}
where the current term should result from the other parts in the
Boltzmann equation and the collision integral is assumed to conserve
the particle number.  We can actually rewrite this into a form that
will turn out to be convenient for later discussions;
\begin{equation}
  \frac{\partial n}{\partial t} + \frac{\partial}{\partial z}
  \biggl(\frac{z}{t}n\biggr) + \bnabla_{\bx_\perp}\cdot \bj_\perp = 0\;,
\label{eq:continuity2}
\end{equation}
for $z\approx 0$.  Here, $z/t$ is a velocity associated with
longitudinal expansion, and thus, $(z/t)n$ gives a longitudinal
current associated with boost invariant expansion.

So far we have considered the situation without electromagnetic fields
for simplicity, the presence of $\bE$ and $\bB$ would result in the
same structure of the continuity equation as
Eq.~\eqref{eq:continuity2}.  In the case of the ordinary Boltzmann
equation with $\bE$ and $\bB$, the Lorentz form terms
$\propto \bE+\bv\times\bB$ are just surface terms and become vanishing
after the momentum integration.

An external electric field should accelerate charged particles, so one
might worry about the fate of boost invariance, though it should exist
formally as argued in Sec.~\ref{sec:Dirac}.  We shall than take a
trivial example to deepen our insight here, that is, the ideal
magneto-hydrodynamics~\cite{Roy:2015kma,Pu:2016ayh}.  By construction
of the \textit{ideal} magneto-hydrodynamics, the electric conductivity
is taken to be infinite, and the electromagnetic force satisfies
$\bE+\bv \times \bB = 0$, which means no force, and thus, no
acceleration of charged particles.  Naturally, the boost invariance is
intact, which addresses another evidence for the existence of boost
invariant states.

\section{Extension to the Chiral Kinetic Theory}
\label{sec:CKT}

We shall treat the CKT under the same Ansatz for the distribution
function with boost invariance.  The CKT for right-handed particles
is~\cite{Stephanov:2012jts,Son:2012wh,Son:2012zy,Chen:2012ca,Hidaka:2016yjf}
\begin{align}
  & (1+\bB\cdot\bOmega) \frac{\partial f}{\partial t} 
  + \bigl[ \bv_p + (\bv_p \cdot \bOmega) \bB
    + \bE \times \bOmega \bigr] \cdot \frac{\partial f}{\partial \bx}
  \notag\\
  & + \bigl[ \bE + \bv_p \times \bB +(\bE \cdot \bB) \bOmega \bigr]
  \cdot \frac{\partial f}{\partial \bp} = I_{\rm coll}[f]\;,
\label{CKE}
\end{align}
where $\bOmega \equiv \hbar\bp /(2|\bp|^3)$ is the Berry curvature,
$\bv_p \equiv \partial\varepsilon_p/\partial\bp$ is the group velocity
with the quasi-particle dispersion relation,
$\varepsilon_p = |\bp|(1-\bB\cdot \bOmega)$ (for the case of spatial
homogeneity), and $I_{\rm coll}[f]$ is the collision integral.  In
this paper we will focus on the $I_{\rm coll}[f]=0$ case, assuming that
the quark distribution is dilute in the early time dynamics of the
heavy-ion collision.

Before we start our discussion, it would be instructive to explain
more about the chiral anomaly and the boost invariance.  As carefully
worked out in Ref.~\cite{Kharzeev:2001ev}, with boost invariance,
there is no large gauge transformation that changes the Chern number
$\nu$ due to trivial homotopy.  Then, if the initial $\nu$ is
vanishing, $\nu$ is always zero, and the winding number,
$Q_w=\nu(\tau=\infty)-\nu(\tau=-\infty)$, is also trivial.  However,
this does not mean that the CKT with boost invariance is trivial.  We
usually consider the CKT for given gauge backgrounds.  One may then
consider some gauge configurations not necessarily given by the
solutions of the equations of motion.  A simple example of constant
(and thus boost invariant) electromagnetic $\bE\cdot\bB$ can actually
make $Q_w$ as large as one likes, irrespective to the homotopy.

As mentioned already in Sec.~\ref{sec:BjorkenBE} we limit ourselves to
deal with a region with $\eta\approx 0$ and $z\approx 0$.  A rather
superficial but controversial difficulty in this case is the
appearance of the Berry curvature that also involves $p_z$ and it
seemingly violates the dependence through $y-\eta$ only.  However, as
argued in the previous sections, the Lorentz invariance immediately
concludes such functional dependence through $y-\eta$;  it would be an
intriguing theoretical problem to confirm this $y-\eta$ dependence
explicitly for arbitrary $\eta$ and $z$.  In this work, we take a more
pragmatic strategy as follows.  We now anticipate that the
distribution function $f$ be a function of $y-\eta$, which is what we
mean by boost invariance;  $f$ itself is invariant under simultaneous
shifts in $y$ and $\eta$.  This requirement would provide us with a
relation like Eq.~\eqref{eq:mid-rapidity}, so that we can discuss a
continuity relation analogously to Eqs.~\eqref{eq:continuity} and
\eqref{eq:continuity2}.

The correct boost transformation onto the CKT is known up to
$\calO(\hbar)$ as $\delta t = \bbeta\cdot\bx$,
$\delta\bx = \bbeta t + \bbeta\times p\bOmega$,
$\delta\bp = \bbeta \varepsilon_p + (\bbeta\times p\bOmega)\times\bB$, 
where $p=|\bp|$~\cite{Chen:2014cla}.
The electromagnetic fields should transform as ordinarily as
$\delta\bB=\bbeta\times\bE$ and $\delta\bE=-\bbeta\times\bB$ up to
this order.

Under our treatment of the problem near $z\approx 0$ we can retain
the leading terms in the expansion of small $\delta z$ around $z=0$ or
small boost parameter $\bbeta=-(\delta z/t)\be_z$.  The boost
transformation then reads,
\begin{align}
  \delta\bx &= -\delta z\,\be_z
  - \frac{\delta z}{t} \be_z\times p\bOmega\;,
  \label{eq:deltax} \\
  \delta\bp &=
    - \delta z \frac{p}{t}\bigl(1-2\bB\cdot\bOmega\bigr)\be_z
    - \delta z \frac{p}{t} B_z \bOmega\;,
  \label{eq:deltap}
\end{align}
up to the linear order in $\delta z$ and in $\bOmega$.  The additional
terms in $\bx$ and $\bp$ proportional to $\bOmega$ are often referred
to as the side-jump terms.  We note that there is a way to formulate
the Lorentz transformation with side-jump terms incorporated not in
$\bx$ and $\bp$ but in $f$ as discussed in
Refs.~\cite{Chen:2015jop,Hidaka:2016yjf}, and we have checked that
both formulations eventually yield the same answer.  The similar
procedures to obtain Eq.~\eqref{eq:mid-rapidity} replace the $z$
dependence with the $p_z$ dependence, which immediately follows from
$f(\bp,\bx,t)=f(\bp+\delta\bp,\bx+\delta\bx,t)$.  The explicit form of
the relation is
\begin{equation}
  \begin{split}
  \label{dfdz}
  \frac{\partial f}{\partial z}
  & = -\frac{B_z p}{t}\bOmega\cdot\frac{\partial f}{\partial\bp}
     - \frac{p}{t}\bigl( 1-2\bB\cdot\bOmega \bigr)
       \frac{\partial f}{\partial p_z} \\
  &\qquad -\frac{p}{t}(\be_z\times\bOmega)_\perp\cdot
   \frac{\partial f}{\partial \bx_\perp}\;.
  \end{split}
\end{equation}
We then plug the above into the original CKT and drop $\calO(\hbar^2)$
terms.  We finally arrive at the following expression;
\begin{align}
  & (1+\bB\cdot\bOmega)\frac{\partial f}{\partial t}
  + \biggl[ ( 1 + 2\bB\cdot\bOmega)\hat{\bp}
  + \bE\times\bOmega \notag\\
  &\quad -\frac{p_z}{t}\, \be_z\times\bOmega \biggr]_\perp\cdot
  \frac{\partial f}{\partial \bx_\perp}
  + \biggl[ \bE+(1+2\bB\cdot\bOmega)\hat{\bp}\times\bB \notag\\
  &\quad +(\bE\cdot\bB)\bOmega -\frac{B_z p_z}{t}\bOmega \notag\\
  &\quad - \frac{p}{t}\bigl[\hat{p}_z + (\bE\times\bOmega)_z \bigr]
  \be_z\biggr] \cdot \frac{\partial f}{\partial \bp}
  = 0\;.
\label{eq:kinetic}
\end{align}
It is easy to confirm that this kinetic equation reduces to the
well-known form of Eq.~\eqref{eq:mid-rapidity} in the $\bOmega\to 0$
limit.  This Eq.~\eqref{eq:kinetic} is the central result in this
paper.

\section{Particle Number Conservation}
\label{sec:number}

It may not be easy to grasp the physical meaning of the
result~\eqref{eq:kinetic}.  To make the physical meaning more
accessible, let us address the continuity equation specifically here,
which is obtained from the zeroth moment of the kinetic equation in
general.  
The definition of the particle number density is slightly modified by 
the Berry curvature correction as~\cite{Xiao:2005eif}
\begin{equation}
  n \equiv \int_{\bp} (1+\bB\cdot\bOmega) f\;,
\label{eq:density}
\end{equation}
and then the momentum integration of Eq.~\eqref{eq:kinetic} leads to
\begin{equation}
  \frac{\partial n}{\partial t} + \frac{\tilde{n}}{t}
  + \bnabla_\perp\cdot\tilde{\bj}_\perp
  = \frac{\bE\cdot\bB}{(2\pi\hbar)^2}f(\bp=0)\;,
\label{eq:cktcontinuity}
\end{equation}
for constant $\bE$ and $\bB$, where the right-hand side is the chiral
anomaly, and we defined the modified density and the modified
transverse current as
\begin{equation}
  \tilde{n} \equiv n - \int_{\bp} \bigl[2\hat{p}_z (\bE\times\bOmega)_z
  +\bB_\perp\cdot\bOmega_\perp \bigr] f\;,
\label{eq:modifieddensity}
\end{equation}
and 
\begin{equation}
  \tilde{\bj}_\perp \equiv \int_{\bp} \Bigl[ (1+2\bB\cdot\bOmega)\hat{\bp}
  + \bE\times\bOmega - \frac{p_z}{t}\,\be_z\times\bOmega\Bigr]_\perp f\;.
\label{eq:j_perp}
\end{equation}
We note that the third term in $\tilde{\bj}_\perp$, i.e.\
$-\int_{\bp} (p_z/t)(\be_z\times\bOmega)_\perp f$, arises from the
side-jump effects in $\bx$ referring to $\delta \bx$ in
Eq.~\eqref{eq:deltax}.  The appearance of such an additional transverse
current is peculiar in the longitudinally expanding situation, while
the other terms in $\tilde{\bj}_\perp$ are just standard ones in the
CKT~\cite{Stephanov:2012jts,Son:2012wh,Son:2012zy,Chen:2012ca,Hidaka:2016yjf}.
We also note that those additional terms in $\tilde{\bj}_\perp$ as
well as those in $\tilde{n}$ are non-vanishing only when $f$ has
an anisotropic distribution in momentum space.

It is useful to write the continuity equation in a way similar to
Eq.~\eqref{eq:continuity2}, that is, from Eq.~\eqref{eq:cktcontinuity}
we have,
\begin{equation}
  \frac{\partial n}{\partial t} 
   + \frac{\partial}{\partial z}\biggl(\frac{z}{t} \tilde{n}\biggl)
   + \bnabla_\perp\cdot \tilde{\bj}_\perp
   = \frac{\bE\cdot\bB}{(2\pi\hbar)^2}f(\bp=0)\;,
\label{eq:cktcontinuity2}
\end{equation}
for $z\approx 0$. Then, $(z/t)\tilde{n}$ represents a current
associated with the longitudinally expanding medium.  Importantly,
here, the density corresponding to this longitudinal current is not
characterized by $n$ itself but $\tilde{n}$, modified by the Berry
curvature as in Eq.~\eqref{eq:modifieddensity}.

\begin{figure}
  \centering
  \includegraphics[bb= 0 0 467 471, width=0.7\columnwidth]{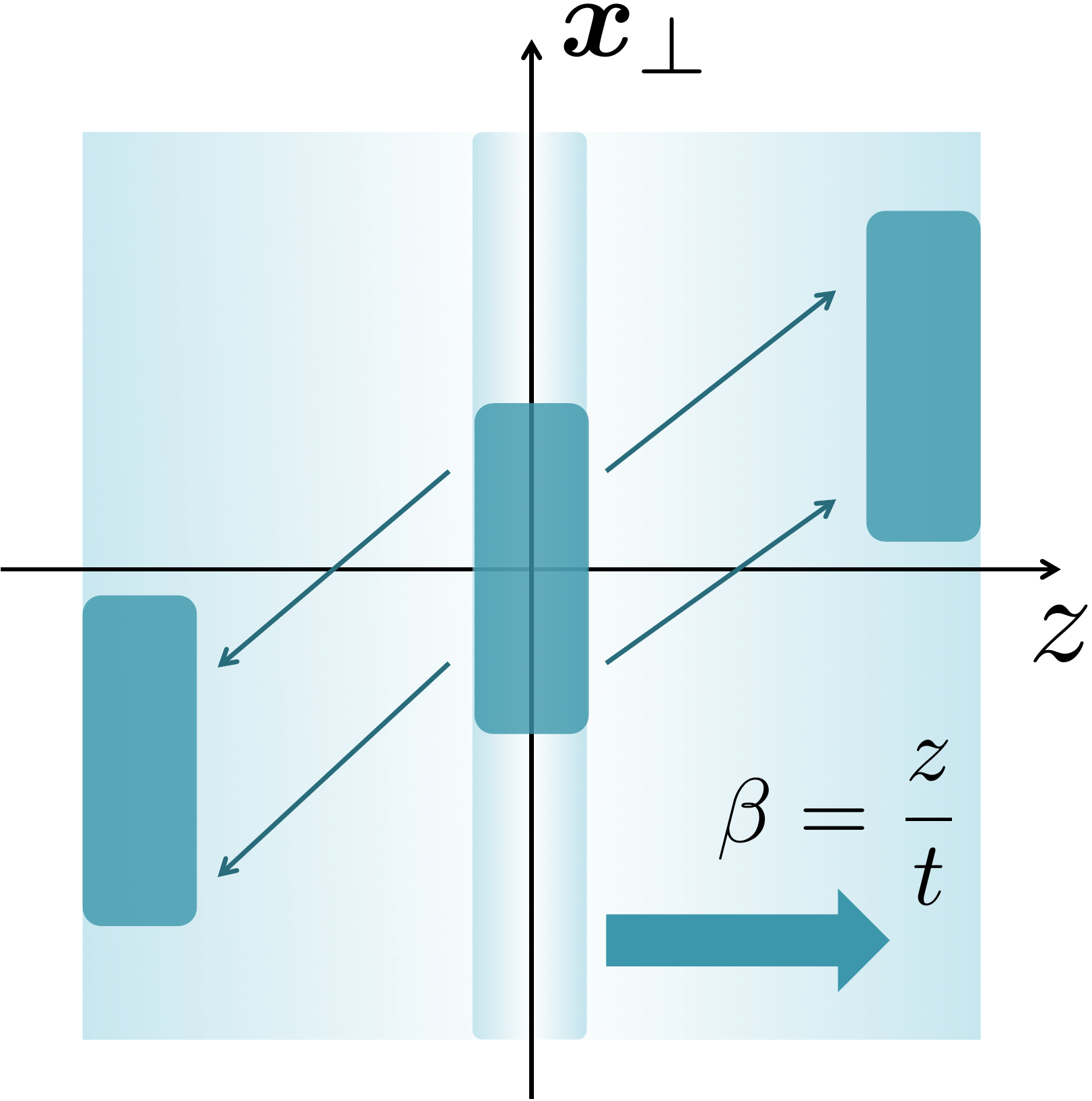}
  \caption{Schematic figure to explain how the shift takes place.}
  \label{fig:expansion}
\end{figure}

\begin{figure}
  \centering
  \includegraphics[bb= 0 0 469 472, width=0.7\columnwidth]{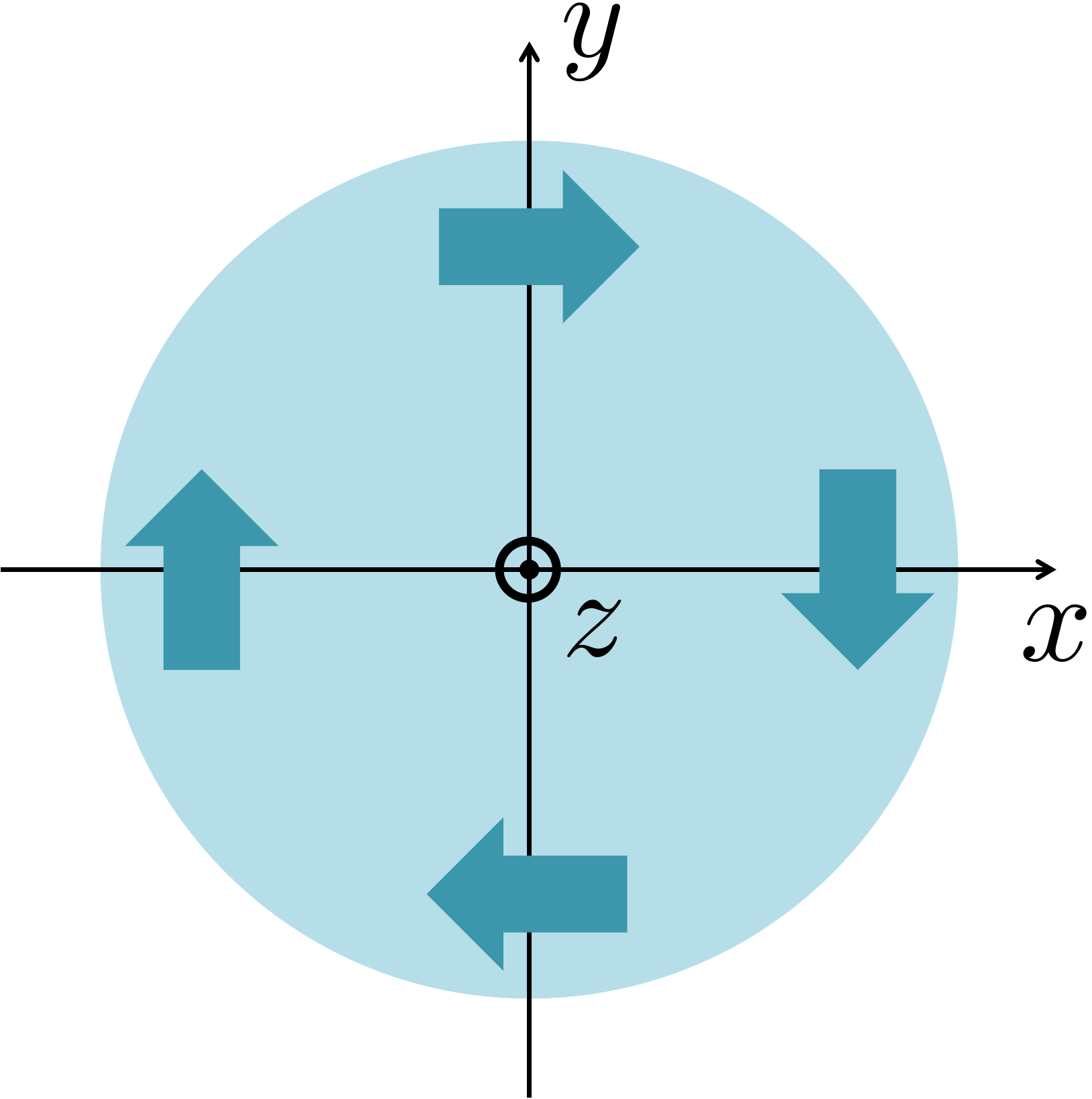}
  \caption{Schematic figure to show the characteristic shift pattern
    seen from the transverse plane for $z>0$.}
  \label{fig:shift}
\end{figure}

Even without external electromagnetic fields, a longitudinal boost
induces the side-jump effects as a result of Eq.~\eqref{eq:deltax}.
Then, apart from the ordinary current, $\int_{\bp}\hat{\bp}_\perp f$, 
only the last term in $\tilde{\bj}_\perp$ given in
Eq.~\eqref{eq:j_perp}, that is,
$-\int_{\bp} (p_z/t)(\be_z\times\bOmega)_\perp f$, remains, which is
absent for the ordinary Boltzmann equation.  We make a schematic
illustration in Fig.~\ref{fig:expansion};  the shift in $\delta\bx$ is
in the direction of $\propto -\text{sgn}(z)\be_z\times \bp$.  Thus, if
the momentum direction is perpendicular to the sheet of
Fig.~\ref{fig:expansion}, the shift goes either upward or downward
depending on the sign of $z$.

We shall point out that the last term in $\tilde{\bj}_\perp$ shows an
intriguing characteristic pattern on the transverse plane.  We can
safely assume that the momentum vector and the coordinate vector are
correlated because the collision point is localized and the system is
expanding.  Then, for example, around the region where $x\gg y$, we
can expect a transverse momentum component along positive $x$ (due to
\textit{transverse} expansion that we have not explicitly considered
in our analysis so far).  Then, $p_z$ is positive for most of particles 
in the region $z>0$ and the
last term in $\tilde{\bj}_\perp$ is directed along
$-\be_z\times\be_x=-\be_y$.  In the same way we can continue our
discussions to conclude that the last term in $\tilde{\bj}_\perp$
exhibits a circular structure as depicted in Fig.~\ref{fig:shift}.
This transverse current, which we may call the
\textit{chiral circular displacement}, is our new finding that appears from
the combination of the boost invariant expansion and the side-jump
effects.  We note that this current is not vector but axial vector
together with the opposite contribution from the left-handed
particles.

The presence of the electromagnetic fields changes nothing in the
above argument;  the CME and the anomalous Hall currents appear then,
but these are already $\mathcal{O}(\hbar)$ effects, and therefore,
should be unaffected by the Berry curvature corrections in the Lorentz
transformation.  So, on the transverse plane, the only remaining shift
effect is the last term in $\tilde{\bj}_\perp$ regardless of the
electromagnetic fields.

The interpretation of $\tilde{n}$ is more complicated.  The difference
between $\tilde{n}$ and $n$ actually comes from two combined effects.
The first one is the additional term $\propto \bOmega$ in $\delta\bp$,
i.e.\ the side-jump effect in momentum according to the requirement of
the boost invariant distribution function.  As seen in
Eq.~\eqref{dfdz}, when we consider $\partial_z{f}$, this momentum
shift is expanded in a form of the momentum derivative of $f$
proportional to $\bOmega$.  After the integration by parts in momentum
space, it is obviously that those terms will become a finite
contribution to $\tilde{n}$.

The second source is the product of the CME and the anomalous Hall
currents, i.e.\ $(\hat{\bp}\cdot\bOmega)\bB+\bE\times\bOmega$, in the
CKT and an ordinary boost term $-(p/t)(\partial f/\partial p_z)$ in
Eq.~\eqref{dfdz} originating from $\delta\bp$.  Because the latter
is a standard term of $\mathcal{O}(1)$, such a combination remains up
to the $\mathcal{O}(\hbar)$ expansion.  Again, after the integration
by parts in momentum space, a finite contribution to $\tilde{n}$
appears.

One might be tempted to given an interpretation to $\tilde{n}$ as an
anomalously induced density, but we must be careful;  the conserved
density is still $n$ as seen in Eqs.~\eqref{eq:cktcontinuity} and
\eqref{eq:cktcontinuity2}, and $\tilde{n}$ is just a ``density'' for
the longitudinal current if it is expressed as a product of the flow
velocity, $z/t$, times the ``density'' that may differ from the
genuine conserved quantity.  Therefore, it is not appropriate to
associate our $\tilde{\bj}_\perp$ and $\tilde{n}$ with any additional
production of particles.

\section{Summary}
\label{summary}

In this paper we have investigated the CKT, the chiral kinetic theory,
with boost invariance.  First, we considered the Dirac equation with
boost invariant external fields.
We made a brief review of the explicit construction of mode functions
as solutions to the Dirac equation and explained that they depend on
$y-\eta$ only, which is the difference between the momentum rapidity
and the coordinate rapidity.  This is the precise meaning of boost
invariance in the wave-functions, while boost invariant external
fields usually refer to $\eta$ independent electromagnetic fields.
This observation is a clear justification for the treatment of the
distribution function $f$ as a function of $y-\eta$.  In fact, in
preceding works, it is a common theoretical approach to utilize the
Boltzmann equation under such an assumption of the function of
$y-\eta$ only to investigate the heavy-ion collision dynamics.

Next, in our main discussions, we extended the idea to the CKT with
boost invariance.  Importantly, the validity region of the kinetic
theory for quark should be significantly wider than that for gluons,
which strongly motivates our present study.
From the explicit solutions of the Dirac equation,
it is guaranteed that the system can still maintain the boost
invariance even though the electromagnetic fields induce the Lorentz
force on the charged particles.
When we boost the system described by the CKT, the non-trivial Lorentz
transformation makes the whole formulation quite complicated as
compared to the conventional Boltzmann case.
Our main result is summarized in Eq.~\eqref{eq:kinetic}.

Then, to develop a more intuitive physical insight, we considered the
continuity equation \eqref{eq:cktcontinuity} from the zeroth moment of the kinetic equation.  In
this way we found that the transverse currents as well as the
longitudinal current take highly non-trivial corrections induced by
the Berry curvature.  In particular, the transverse currents have
non-zero corrections even for vanishing external fields, which are
purely induced by the side-jump effects in the Lorentz transformation
for coordinates.  We pointed out that these corrections show a
characteristic distribution pattern, which we named the
\textit{chiral circular displacement}.  It is an interesting possibility
that the chiral circular displacement may cause some spatial distribution
with chiral imbalance for non-central heavy-ion collisions.  For more
detailed discussions on phenomenological implications, we will report
separately, and we would make this paper focused on the formal aspect
of the boost invariant CKT formulation.

It is still a big theoretical challenge to apply the CKT to the full
simulations for heavy-ion phenomenology including the expansion
effects.
Even though we argued that we can neglect the interactions among
quarks due to diluteness in the initial state, we must eventually take
account of the interaction effect and consider the collision integral.
Then, the treatment of the CKT would become even more complicated due
to the side-jump effects in the interactions, which could be
incorporated in the so-called no-jump frame~\cite{Chen:2015jop,Hidaka:2016yjf}.
We need more theoretical
developments along these lines.

\begin{acknowledgments}
 We thank Yoshimasa~Hidaka, Dimtri~Kharzeev, Naoto~Tanji, Yi~Yin, and
Di-Lun~Yang for helpful discussions. S.~P.\ expresses a special thanks
to organizers of the Chirality QCD Workshop 2017 for helpful
discussions.
S.~E.\ was supported by Grant-in-Aid for JSPS Fellows Grant 
No.\ 17J02380.
K.~F.\ was supported by JSPS KAKENHI Grant
No.\ 15H03652 and 15K13479.
S.~P.\ was supported by JSPS 
post-doctoral fellowship for foreign researchers. 

\end{acknowledgments}

\bibliography{expansion}

\begin{thebibliography}{66}%
\makeatletter
\providecommand \@ifxundefined [1]{%
 \@ifx{#1\undefined}
}%
\providecommand \@ifnum [1]{%
 \ifnum #1\expandafter \@firstoftwo
 \else \expandafter \@secondoftwo
 \fi
}%
\providecommand \@ifx [1]{%
 \ifx #1\expandafter \@firstoftwo
 \else \expandafter \@secondoftwo
 \fi
}%
\providecommand \natexlab [1]{#1}%
\providecommand \enquote  [1]{``#1''}%
\providecommand \bibnamefont  [1]{#1}%
\providecommand \bibfnamefont [1]{#1}%
\providecommand \citenamefont [1]{#1}%
\providecommand \href@noop [0]{\@secondoftwo}%
\providecommand \href [0]{\begingroup \@sanitize@url \@href}%
\providecommand \@href[1]{\@@startlink{#1}\@@href}%
\providecommand \@@href[1]{\endgroup#1\@@endlink}%
\providecommand \@sanitize@url [0]{\catcode `\\12\catcode `\$12\catcode
  `\&12\catcode `\#12\catcode `\^12\catcode `\_12\catcode `\%12\relax}%
\providecommand \@@startlink[1]{}%
\providecommand \@@endlink[0]{}%
\providecommand \url  [0]{\begingroup\@sanitize@url \@url }%
\providecommand \@url [1]{\endgroup\@href {#1}{\urlprefix }}%
\providecommand \urlprefix  [0]{URL }%
\providecommand \Eprint [0]{\href }%
\providecommand \doibase [0]{http://dx.doi.org/}%
\providecommand \selectlanguage [0]{\@gobble}%
\providecommand \bibinfo  [0]{\@secondoftwo}%
\providecommand \bibfield  [0]{\@secondoftwo}%
\providecommand \translation [1]{[#1]}%
\providecommand \BibitemOpen [0]{}%
\providecommand \bibitemStop [0]{}%
\providecommand \bibitemNoStop [0]{.\EOS\space}%
\providecommand \EOS [0]{\spacefactor3000\relax}%
\providecommand \BibitemShut  [1]{\csname bibitem#1\endcsname}%
\let\auto@bib@innerbib\@empty
\bibitem [{\citenamefont {Kharzeev}\ \emph {et~al.}(2008)\citenamefont
  {Kharzeev}, \citenamefont {McLerran},\ and\ \citenamefont
  {Warringa}}]{Kharzeev:2007jp}%
  \BibitemOpen
  \bibfield  {author} {\bibinfo {author} {\bibfnamefont {D.~E.}\ \bibnamefont
  {Kharzeev}}, \bibinfo {author} {\bibfnamefont {L.~D.}\ \bibnamefont
  {McLerran}}, \ and\ \bibinfo {author} {\bibfnamefont {H.~J.}\ \bibnamefont
  {Warringa}},\ }\href {\doibase 10.1016/j.nuclphysa.2008.02.298} {\bibfield
  {journal} {\bibinfo  {journal} {Nucl. Phys.}\ }\textbf {\bibinfo {volume}
  {A803}},\ \bibinfo {pages} {227} (\bibinfo {year} {2008})},\ \Eprint
  {http://arxiv.org/abs/0711.0950} {arXiv:0711.0950 [hep-ph]} \BibitemShut
  {NoStop}%
\bibitem [{\citenamefont {Fukushima}\ \emph {et~al.}(2008)\citenamefont
  {Fukushima}, \citenamefont {Kharzeev},\ and\ \citenamefont
  {Warringa}}]{Fukushima:2008xe}%
  \BibitemOpen
  \bibfield  {author} {\bibinfo {author} {\bibfnamefont {K.}~\bibnamefont
  {Fukushima}}, \bibinfo {author} {\bibfnamefont {D.~E.}\ \bibnamefont
  {Kharzeev}}, \ and\ \bibinfo {author} {\bibfnamefont {H.~J.}\ \bibnamefont
  {Warringa}},\ }\href {\doibase 10.1103/PhysRevD.78.074033} {\bibfield
  {journal} {\bibinfo  {journal} {Phys. Rev.}\ }\textbf {\bibinfo {volume}
  {D78}},\ \bibinfo {pages} {074033} (\bibinfo {year} {2008})},\ \Eprint
  {http://arxiv.org/abs/0808.3382} {arXiv:0808.3382 [hep-ph]} \BibitemShut
  {NoStop}%
\bibitem [{\citenamefont {Kharzeev}\ \emph {et~al.}(2016)\citenamefont
  {Kharzeev}, \citenamefont {Liao}, \citenamefont {Voloshin},\ and\
  \citenamefont {Wang}}]{Kharzeev:2015znc}%
  \BibitemOpen
  \bibfield  {author} {\bibinfo {author} {\bibfnamefont {D.~E.}\ \bibnamefont
  {Kharzeev}}, \bibinfo {author} {\bibfnamefont {J.}~\bibnamefont {Liao}},
  \bibinfo {author} {\bibfnamefont {S.~A.}\ \bibnamefont {Voloshin}}, \ and\
  \bibinfo {author} {\bibfnamefont {G.}~\bibnamefont {Wang}},\ }\href {\doibase
  10.1016/j.ppnp.2016.01.001} {\bibfield  {journal} {\bibinfo  {journal} {Prog.
  Part. Nucl. Phys.}\ }\textbf {\bibinfo {volume} {88}},\ \bibinfo {pages} {1}
  (\bibinfo {year} {2016})},\ \Eprint {http://arxiv.org/abs/1511.04050}
  {arXiv:1511.04050 [hep-ph]} \BibitemShut {NoStop}%
\bibitem [{\citenamefont {Hattori}\ and\ \citenamefont
  {Huang}(2017)}]{Hattori:2016emy}%
  \BibitemOpen
  \bibfield  {author} {\bibinfo {author} {\bibfnamefont {K.}~\bibnamefont
  {Hattori}}\ and\ \bibinfo {author} {\bibfnamefont {X.-G.}\ \bibnamefont
  {Huang}},\ }\href {\doibase 10.1007/s41365-016-0178-3} {\bibfield  {journal}
  {\bibinfo  {journal} {Nucl. Sci. Tech.}\ }\textbf {\bibinfo {volume} {28}},\
  \bibinfo {pages} {26} (\bibinfo {year} {2017})},\ \Eprint
  {http://arxiv.org/abs/1609.00747} {arXiv:1609.00747 [nucl-th]} \BibitemShut
  {NoStop}%
\bibitem [{\citenamefont {Son}\ and\ \citenamefont
  {Surowka}(2009)}]{Son:2009tf}%
  \BibitemOpen
  \bibfield  {author} {\bibinfo {author} {\bibfnamefont {D.~T.}\ \bibnamefont
  {Son}}\ and\ \bibinfo {author} {\bibfnamefont {P.}~\bibnamefont {Surowka}},\
  }\href {\doibase 10.1103/PhysRevLett.103.191601} {\bibfield  {journal}
  {\bibinfo  {journal} {Phys. Rev. Lett.}\ }\textbf {\bibinfo {volume} {103}},\
  \bibinfo {pages} {191601} (\bibinfo {year} {2009})},\ \Eprint
  {http://arxiv.org/abs/0906.5044} {arXiv:0906.5044 [hep-th]} \BibitemShut
  {NoStop}%
\bibitem [{\citenamefont {Sadofyev}\ and\ \citenamefont
  {Isachenkov}(2011)}]{Sadofyev:2010pr}%
  \BibitemOpen
  \bibfield  {author} {\bibinfo {author} {\bibfnamefont {A.~V.}\ \bibnamefont
  {Sadofyev}}\ and\ \bibinfo {author} {\bibfnamefont {M.~V.}\ \bibnamefont
  {Isachenkov}},\ }\href {\doibase 10.1016/j.physletb.2011.02.041} {\bibfield
  {journal} {\bibinfo  {journal} {Phys. Lett.}\ }\textbf {\bibinfo {volume}
  {B697}},\ \bibinfo {pages} {404} (\bibinfo {year} {2011})},\ \Eprint
  {http://arxiv.org/abs/1010.1550} {arXiv:1010.1550 [hep-th]} \BibitemShut
  {NoStop}%
\bibitem [{\citenamefont {Pu}\ \emph {et~al.}(2011)\citenamefont {Pu},
  \citenamefont {Gao},\ and\ \citenamefont {Wang}}]{Pu:2010as}%
  \BibitemOpen
  \bibfield  {author} {\bibinfo {author} {\bibfnamefont {S.}~\bibnamefont
  {Pu}}, \bibinfo {author} {\bibfnamefont {J.-h.}\ \bibnamefont {Gao}}, \ and\
  \bibinfo {author} {\bibfnamefont {Q.}~\bibnamefont {Wang}},\ }\href {\doibase
  10.1103/PhysRevD.83.094017} {\bibfield  {journal} {\bibinfo  {journal} {Phys.
  Rev.}\ }\textbf {\bibinfo {volume} {D83}},\ \bibinfo {pages} {094017}
  (\bibinfo {year} {2011})},\ \Eprint {http://arxiv.org/abs/1008.2418}
  {arXiv:1008.2418 [nucl-th]} \BibitemShut {NoStop}%
\bibitem [{\citenamefont {Abramczyk}\ \emph {et~al.}(2009)\citenamefont
  {Abramczyk}, \citenamefont {Blum}, \citenamefont {Petropoulos},\ and\
  \citenamefont {Zhou}}]{Abramczyk:2009gb}%
  \BibitemOpen
  \bibfield  {author} {\bibinfo {author} {\bibfnamefont {M.}~\bibnamefont
  {Abramczyk}}, \bibinfo {author} {\bibfnamefont {T.}~\bibnamefont {Blum}},
  \bibinfo {author} {\bibfnamefont {G.}~\bibnamefont {Petropoulos}}, \ and\
  \bibinfo {author} {\bibfnamefont {R.}~\bibnamefont {Zhou}},\ }\href@noop {}
  {\bibfield  {journal} {\bibinfo  {journal} {PoS}\ }\textbf {\bibinfo {volume}
  {LAT2009}},\ \bibinfo {pages} {181} (\bibinfo {year} {2009})},\ \Eprint
  {http://arxiv.org/abs/0911.1348} {arXiv:0911.1348 [hep-lat]} \BibitemShut
  {NoStop}%
\bibitem [{\citenamefont {Buividovich}\ \emph
  {et~al.}(2009{\natexlab{a}})\citenamefont {Buividovich}, \citenamefont
  {Chernodub}, \citenamefont {Luschevskaya},\ and\ \citenamefont
  {Polikarpov}}]{Buividovich:2009wi}%
  \BibitemOpen
  \bibfield  {author} {\bibinfo {author} {\bibfnamefont {P.}~\bibnamefont
  {Buividovich}}, \bibinfo {author} {\bibfnamefont {M.}~\bibnamefont
  {Chernodub}}, \bibinfo {author} {\bibfnamefont {E.}~\bibnamefont
  {Luschevskaya}}, \ and\ \bibinfo {author} {\bibfnamefont {M.}~\bibnamefont
  {Polikarpov}},\ }\href {\doibase 10.1103/PhysRevD.80.054503} {\bibfield
  {journal} {\bibinfo  {journal} {Phys. Rev.}\ }\textbf {\bibinfo {volume}
  {D80}},\ \bibinfo {pages} {054503} (\bibinfo {year} {2009}{\natexlab{a}})},\
  \Eprint {http://arxiv.org/abs/0907.0494} {arXiv:0907.0494 [hep-lat]}
  \BibitemShut {NoStop}%
\bibitem [{\citenamefont {Buividovich}\ \emph
  {et~al.}(2009{\natexlab{b}})\citenamefont {Buividovich}, \citenamefont
  {Luschevskaya}, \citenamefont {Polikarpov},\ and\ \citenamefont
  {Chernodub}}]{Buividovich:2009zzb}%
  \BibitemOpen
  \bibfield  {author} {\bibinfo {author} {\bibfnamefont {P.}~\bibnamefont
  {Buividovich}}, \bibinfo {author} {\bibfnamefont {E.}~\bibnamefont
  {Luschevskaya}}, \bibinfo {author} {\bibfnamefont {M.}~\bibnamefont
  {Polikarpov}}, \ and\ \bibinfo {author} {\bibfnamefont {M.}~\bibnamefont
  {Chernodub}},\ }\href {\doibase 10.1134/S0021364009180027} {\bibfield
  {journal} {\bibinfo  {journal} {JETP Lett.}\ }\textbf {\bibinfo {volume}
  {90}},\ \bibinfo {pages} {412} (\bibinfo {year}
  {2009}{\natexlab{b}})}\BibitemShut {NoStop}%
\bibitem [{\citenamefont {Buividovich}\ \emph {et~al.}(2010)\citenamefont
  {Buividovich}, \citenamefont {Chernodub}, \citenamefont {Kharzeev},
  \citenamefont {Kalaydzhyan}, \citenamefont {Luschevskaya} \emph
  {et~al.}}]{Buividovich:2010tn}%
  \BibitemOpen
  \bibfield  {author} {\bibinfo {author} {\bibfnamefont {P.}~\bibnamefont
  {Buividovich}}, \bibinfo {author} {\bibfnamefont {M.}~\bibnamefont
  {Chernodub}}, \bibinfo {author} {\bibfnamefont {D.}~\bibnamefont {Kharzeev}},
  \bibinfo {author} {\bibfnamefont {T.}~\bibnamefont {Kalaydzhyan}}, \bibinfo
  {author} {\bibfnamefont {E.}~\bibnamefont {Luschevskaya}},  \emph {et~al.},\
  }\href {\doibase 10.1103/PhysRevLett.105.132001} {\bibfield  {journal}
  {\bibinfo  {journal} {Phys. Rev. Lett.}\ }\textbf {\bibinfo {volume} {105}},\
  \bibinfo {pages} {132001} (\bibinfo {year} {2010})},\ \Eprint
  {http://arxiv.org/abs/1003.2180} {arXiv:1003.2180 [hep-lat]} \BibitemShut
  {NoStop}%
\bibitem [{\citenamefont {Yamamoto}(2011)}]{Yamamoto:2011gk}%
  \BibitemOpen
  \bibfield  {author} {\bibinfo {author} {\bibfnamefont {A.}~\bibnamefont
  {Yamamoto}},\ }\href {\doibase 10.1103/PhysRevLett.107.031601} {\bibfield
  {journal} {\bibinfo  {journal} {Phys. Rev. Lett.}\ }\textbf {\bibinfo
  {volume} {107}},\ \bibinfo {pages} {031601} (\bibinfo {year} {2011})},\
  \Eprint {http://arxiv.org/abs/1105.0385} {arXiv:1105.0385 [hep-lat]}
  \BibitemShut {NoStop}%
\bibitem [{\citenamefont {Erdmenger}\ \emph {et~al.}(2009)\citenamefont
  {Erdmenger}, \citenamefont {Haack}, \citenamefont {Kaminski},\ and\
  \citenamefont {Yarom}}]{Erdmenger2009}%
  \BibitemOpen
  \bibfield  {author} {\bibinfo {author} {\bibfnamefont {J.}~\bibnamefont
  {Erdmenger}}, \bibinfo {author} {\bibfnamefont {M.}~\bibnamefont {Haack}},
  \bibinfo {author} {\bibfnamefont {M.}~\bibnamefont {Kaminski}}, \ and\
  \bibinfo {author} {\bibfnamefont {A.}~\bibnamefont {Yarom}},\ }\href
  {\doibase 10.1088/1126-6708/2009/01/055} {\bibfield  {journal} {\bibinfo
  {journal} {JHEP}\ }\textbf {\bibinfo {volume} {01}},\ \bibinfo {pages} {055}
  (\bibinfo {year} {2009})},\ \Eprint {http://arxiv.org/abs/0809.2488}
  {arXiv:0809.2488 [hep-th]} \BibitemShut {NoStop}%
\bibitem [{\citenamefont {Torabian}\ and\ \citenamefont
  {Yee}(2009)}]{Torabian2009a}%
  \BibitemOpen
  \bibfield  {author} {\bibinfo {author} {\bibfnamefont {M.}~\bibnamefont
  {Torabian}}\ and\ \bibinfo {author} {\bibfnamefont {H.-U.}\ \bibnamefont
  {Yee}},\ }\href {\doibase 10.1088/1126-6708/2009/08/020} {\bibfield
  {journal} {\bibinfo  {journal} {JHEP}\ }\textbf {\bibinfo {volume} {08}},\
  \bibinfo {pages} {020} (\bibinfo {year} {2009})},\ \Eprint
  {http://arxiv.org/abs/0903.4894} {arXiv:0903.4894 [hep-th]} \BibitemShut
  {NoStop}%
\bibitem [{\citenamefont {Banerjee}\ \emph {et~al.}(2011)\citenamefont
  {Banerjee}, \citenamefont {Bhattacharya}, \citenamefont {Bhattacharyya},
  \citenamefont {Dutta}, \citenamefont {Loganayagam} \emph
  {et~al.}}]{Banerjee2011}%
  \BibitemOpen
  \bibfield  {author} {\bibinfo {author} {\bibfnamefont {N.}~\bibnamefont
  {Banerjee}}, \bibinfo {author} {\bibfnamefont {J.}~\bibnamefont
  {Bhattacharya}}, \bibinfo {author} {\bibfnamefont {S.}~\bibnamefont
  {Bhattacharyya}}, \bibinfo {author} {\bibfnamefont {S.}~\bibnamefont
  {Dutta}}, \bibinfo {author} {\bibfnamefont {R.}~\bibnamefont {Loganayagam}},
  \emph {et~al.},\ }\href {\doibase 10.1007/JHEP01(2011)094} {\bibfield
  {journal} {\bibinfo  {journal} {JHEP}\ }\textbf {\bibinfo {volume} {1101}},\
  \bibinfo {pages} {094} (\bibinfo {year} {2011})},\ \Eprint
  {http://arxiv.org/abs/0809.2596} {arXiv:0809.2596 [hep-th]} \BibitemShut
  {NoStop}%
\bibitem [{\citenamefont {Gao}\ \emph {et~al.}(2012)\citenamefont {Gao},
  \citenamefont {Liang}, \citenamefont {Pu}, \citenamefont {Wang},\ and\
  \citenamefont {Wang}}]{Gao:2012ix}%
  \BibitemOpen
  \bibfield  {author} {\bibinfo {author} {\bibfnamefont {J.-H.}\ \bibnamefont
  {Gao}}, \bibinfo {author} {\bibfnamefont {Z.-T.}\ \bibnamefont {Liang}},
  \bibinfo {author} {\bibfnamefont {S.}~\bibnamefont {Pu}}, \bibinfo {author}
  {\bibfnamefont {Q.}~\bibnamefont {Wang}}, \ and\ \bibinfo {author}
  {\bibfnamefont {X.-N.}\ \bibnamefont {Wang}},\ }\href {\doibase
  10.1103/PhysRevLett.109.232301} {\bibfield  {journal} {\bibinfo  {journal}
  {Phys. Rev. Lett.}\ }\textbf {\bibinfo {volume} {109}},\ \bibinfo {pages}
  {232301} (\bibinfo {year} {2012})},\ \Eprint {http://arxiv.org/abs/1203.0725}
  {arXiv:1203.0725 [hep-ph]} \BibitemShut {NoStop}%
\bibitem [{\citenamefont {Gao}\ \emph {et~al.}(2017)\citenamefont {Gao},
  \citenamefont {Pu},\ and\ \citenamefont {Wang}}]{Gao:2017rgi}%
  \BibitemOpen
  \bibfield  {author} {\bibinfo {author} {\bibfnamefont {J.-h.}\ \bibnamefont
  {Gao}}, \bibinfo {author} {\bibfnamefont {S.}~\bibnamefont {Pu}}, \ and\
  \bibinfo {author} {\bibfnamefont {Q.}~\bibnamefont {Wang}},\ }\href@noop {}
  {\  (\bibinfo {year} {2017})},\ \Eprint {http://arxiv.org/abs/1704.00244}
  {arXiv:1704.00244 [nucl-th]} \BibitemShut {NoStop}%
\bibitem [{\citenamefont {Son}\ and\ \citenamefont
  {Yamamoto}(2012)}]{Son:2012wh}%
  \BibitemOpen
  \bibfield  {author} {\bibinfo {author} {\bibfnamefont {D.~T.}\ \bibnamefont
  {Son}}\ and\ \bibinfo {author} {\bibfnamefont {N.}~\bibnamefont {Yamamoto}},\
  }\href {\doibase 10.1103/PhysRevLett.109.181602} {\bibfield  {journal}
  {\bibinfo  {journal} {Phys. Rev. Lett.}\ }\textbf {\bibinfo {volume} {109}},\
  \bibinfo {pages} {181602} (\bibinfo {year} {2012})},\ \Eprint
  {http://arxiv.org/abs/1203.2697} {arXiv:1203.2697 [cond-mat.mes-hall]}
  \BibitemShut {NoStop}%
\bibitem [{\citenamefont {Son}\ and\ \citenamefont
  {Yamamoto}(2013)}]{Son:2012zy}%
  \BibitemOpen
  \bibfield  {author} {\bibinfo {author} {\bibfnamefont {D.~T.}\ \bibnamefont
  {Son}}\ and\ \bibinfo {author} {\bibfnamefont {N.}~\bibnamefont {Yamamoto}},\
  }\href {\doibase 10.1103/PhysRevD.87.085016} {\bibfield  {journal} {\bibinfo
  {journal} {Phys. Rev.}\ }\textbf {\bibinfo {volume} {D87}},\ \bibinfo {pages}
  {085016} (\bibinfo {year} {2013})},\ \Eprint {http://arxiv.org/abs/1210.8158}
  {arXiv:1210.8158 [hep-th]} \BibitemShut {NoStop}%
\bibitem [{\citenamefont {{Stephanov}}\ and\ \citenamefont
  {{Yin}}(2012)}]{Stephanov:2012jts}%
  \BibitemOpen
  \bibfield  {author} {\bibinfo {author} {\bibfnamefont {M.~A.}\ \bibnamefont
  {{Stephanov}}}\ and\ \bibinfo {author} {\bibfnamefont {Y.}~\bibnamefont
  {{Yin}}},\ }\href {\doibase 10.1103/PhysRevLett.109.162001} {\bibfield
  {journal} {\bibinfo  {journal} {Physical Review Letters}\ }\textbf {\bibinfo
  {volume} {109}},\ \bibinfo {eid} {162001} (\bibinfo {year} {2012})},\ \Eprint
  {http://arxiv.org/abs/1207.0747} {arXiv:1207.0747 [hep-th]} \BibitemShut
  {NoStop}%
\bibitem [{\citenamefont {Chen}\ \emph {et~al.}(2013)\citenamefont {Chen},
  \citenamefont {Pu}, \citenamefont {Wang},\ and\ \citenamefont
  {Wang}}]{Chen:2012ca}%
  \BibitemOpen
  \bibfield  {author} {\bibinfo {author} {\bibfnamefont {J.-W.}\ \bibnamefont
  {Chen}}, \bibinfo {author} {\bibfnamefont {S.}~\bibnamefont {Pu}}, \bibinfo
  {author} {\bibfnamefont {Q.}~\bibnamefont {Wang}}, \ and\ \bibinfo {author}
  {\bibfnamefont {X.-N.}\ \bibnamefont {Wang}},\ }\href {\doibase
  10.1103/PhysRevLett.110.262301} {\bibfield  {journal} {\bibinfo  {journal}
  {Phys. Rev. Lett.}\ }\textbf {\bibinfo {volume} {110}},\ \bibinfo {pages}
  {262301} (\bibinfo {year} {2013})},\ \Eprint {http://arxiv.org/abs/1210.8312}
  {arXiv:1210.8312 [hep-th]} \BibitemShut {NoStop}%
\bibitem [{\citenamefont {Skokov}\ \emph {et~al.}(2009)\citenamefont {Skokov},
  \citenamefont {Illarionov},\ and\ \citenamefont {Toneev}}]{Skokov:2009qp}%
  \BibitemOpen
  \bibfield  {author} {\bibinfo {author} {\bibfnamefont {V.}~\bibnamefont
  {Skokov}}, \bibinfo {author} {\bibfnamefont {A.~{\relax Yu}.}\ \bibnamefont
  {Illarionov}}, \ and\ \bibinfo {author} {\bibfnamefont {V.}~\bibnamefont
  {Toneev}},\ }\href {\doibase 10.1142/S0217751X09047570} {\bibfield  {journal}
  {\bibinfo  {journal} {Int. J. Mod. Phys.}\ }\textbf {\bibinfo {volume}
  {A24}},\ \bibinfo {pages} {5925} (\bibinfo {year} {2009})},\ \Eprint
  {http://arxiv.org/abs/0907.1396} {arXiv:0907.1396 [nucl-th]} \BibitemShut
  {NoStop}%
\bibitem [{\citenamefont {Bzdak}\ and\ \citenamefont
  {Skokov}(2012)}]{Bzdak:2011yy}%
  \BibitemOpen
  \bibfield  {author} {\bibinfo {author} {\bibfnamefont {A.}~\bibnamefont
  {Bzdak}}\ and\ \bibinfo {author} {\bibfnamefont {V.}~\bibnamefont {Skokov}},\
  }\href {\doibase 10.1016/j.physletb.2012.02.065} {\bibfield  {journal}
  {\bibinfo  {journal} {Phys. Lett.}\ }\textbf {\bibinfo {volume} {B710}},\
  \bibinfo {pages} {171} (\bibinfo {year} {2012})},\ \Eprint
  {http://arxiv.org/abs/1111.1949} {arXiv:1111.1949 [hep-ph]} \BibitemShut
  {NoStop}%
\bibitem [{\citenamefont {Deng}\ and\ \citenamefont
  {Huang}(2012)}]{Deng:2012pc}%
  \BibitemOpen
  \bibfield  {author} {\bibinfo {author} {\bibfnamefont {W.-T.}\ \bibnamefont
  {Deng}}\ and\ \bibinfo {author} {\bibfnamefont {X.-G.}\ \bibnamefont
  {Huang}},\ }\href {\doibase 10.1103/PhysRevC.85.044907} {\bibfield  {journal}
  {\bibinfo  {journal} {Phys. Rev.}\ }\textbf {\bibinfo {volume} {C85}},\
  \bibinfo {pages} {044907} (\bibinfo {year} {2012})},\ \Eprint
  {http://arxiv.org/abs/1201.5108} {arXiv:1201.5108 [nucl-th]} \BibitemShut
  {NoStop}%
\bibitem [{\citenamefont {Roy}\ and\ \citenamefont {Pu}(2015)}]{Roy:2015coa}%
  \BibitemOpen
  \bibfield  {author} {\bibinfo {author} {\bibfnamefont {V.}~\bibnamefont
  {Roy}}\ and\ \bibinfo {author} {\bibfnamefont {S.}~\bibnamefont {Pu}},\
  }\href {\doibase 10.1103/PhysRevC.92.064902} {\bibfield  {journal} {\bibinfo
  {journal} {Phys. Rev.}\ }\textbf {\bibinfo {volume} {C92}},\ \bibinfo {pages}
  {064902} (\bibinfo {year} {2015})},\ \Eprint
  {http://arxiv.org/abs/1508.03761} {arXiv:1508.03761 [nucl-th]} \BibitemShut
  {NoStop}%
\bibitem [{\citenamefont {Tuchin}(2015)}]{Tuchin:2014iua}%
  \BibitemOpen
  \bibfield  {author} {\bibinfo {author} {\bibfnamefont {K.}~\bibnamefont
  {Tuchin}},\ }\href {\doibase 10.1103/PhysRevC.91.064902} {\bibfield
  {journal} {\bibinfo  {journal} {Phys. Rev.}\ }\textbf {\bibinfo {volume}
  {C91}},\ \bibinfo {pages} {064902} (\bibinfo {year} {2015})},\ \Eprint
  {http://arxiv.org/abs/1411.1363} {arXiv:1411.1363 [hep-ph]} \BibitemShut
  {NoStop}%
\bibitem [{\citenamefont {Li}\ \emph {et~al.}(2016)\citenamefont {Li},
  \citenamefont {Sheng},\ and\ \citenamefont {Wang}}]{Li:2016tel}%
  \BibitemOpen
  \bibfield  {author} {\bibinfo {author} {\bibfnamefont {H.}~\bibnamefont
  {Li}}, \bibinfo {author} {\bibfnamefont {X.-l.}\ \bibnamefont {Sheng}}, \
  and\ \bibinfo {author} {\bibfnamefont {Q.}~\bibnamefont {Wang}},\ }\href
  {\doibase 10.1103/PhysRevC.94.044903} {\bibfield  {journal} {\bibinfo
  {journal} {Phys. Rev.}\ }\textbf {\bibinfo {volume} {C94}},\ \bibinfo {pages}
  {044903} (\bibinfo {year} {2016})},\ \Eprint
  {http://arxiv.org/abs/1602.02223} {arXiv:1602.02223 [nucl-th]} \BibitemShut
  {NoStop}%
\bibitem [{\citenamefont {Abelev}\ \emph {et~al.}(2009)\citenamefont {Abelev}
  \emph {et~al.}}]{Abelev:2009ac}%
  \BibitemOpen
  \bibfield  {author} {\bibinfo {author} {\bibfnamefont {B.~I.}\ \bibnamefont
  {Abelev}} \emph {et~al.} (\bibinfo {collaboration} {STAR}),\ }\href {\doibase
  10.1103/PhysRevLett.103.251601} {\bibfield  {journal} {\bibinfo  {journal}
  {Phys. Rev. Lett.}\ }\textbf {\bibinfo {volume} {103}},\ \bibinfo {pages}
  {251601} (\bibinfo {year} {2009})},\ \Eprint {http://arxiv.org/abs/0909.1739}
  {arXiv:0909.1739 [nucl-ex]} \BibitemShut {NoStop}%
\bibitem [{\citenamefont {Abelev}\ \emph {et~al.}(2010)\citenamefont {Abelev}
  \emph {et~al.}}]{Abelev:2009ad}%
  \BibitemOpen
  \bibfield  {author} {\bibinfo {author} {\bibfnamefont {B.~I.}\ \bibnamefont
  {Abelev}} \emph {et~al.} (\bibinfo {collaboration} {STAR}),\ }\href {\doibase
  10.1103/PhysRevC.81.054908} {\bibfield  {journal} {\bibinfo  {journal} {Phys.
  Rev.}\ }\textbf {\bibinfo {volume} {C81}},\ \bibinfo {pages} {054908}
  (\bibinfo {year} {2010})},\ \Eprint {http://arxiv.org/abs/0909.1717}
  {arXiv:0909.1717 [nucl-ex]} \BibitemShut {NoStop}%
\bibitem [{\citenamefont {Huang}\ and\ \citenamefont
  {Liao}(2013)}]{Huang:2013iia}%
  \BibitemOpen
  \bibfield  {author} {\bibinfo {author} {\bibfnamefont {X.-G.}\ \bibnamefont
  {Huang}}\ and\ \bibinfo {author} {\bibfnamefont {J.}~\bibnamefont {Liao}},\
  }\href {\doibase 10.1103/PhysRevLett.110.232302} {\bibfield  {journal}
  {\bibinfo  {journal} {Phys. Rev. Lett.}\ }\textbf {\bibinfo {volume} {110}},\
  \bibinfo {pages} {232302} (\bibinfo {year} {2013})},\ \Eprint
  {http://arxiv.org/abs/1303.7192} {arXiv:1303.7192 [nucl-th]} \BibitemShut
  {NoStop}%
\bibitem [{\citenamefont {Pu}\ \emph {et~al.}(2014)\citenamefont {Pu},
  \citenamefont {Wu},\ and\ \citenamefont {Yang}}]{Pu:2014cwa}%
  \BibitemOpen
  \bibfield  {author} {\bibinfo {author} {\bibfnamefont {S.}~\bibnamefont
  {Pu}}, \bibinfo {author} {\bibfnamefont {S.-Y.}\ \bibnamefont {Wu}}, \ and\
  \bibinfo {author} {\bibfnamefont {D.-L.}\ \bibnamefont {Yang}},\ }\href
  {\doibase 10.1103/PhysRevD.89.085024} {\bibfield  {journal} {\bibinfo
  {journal} {Phys. Rev.}\ }\textbf {\bibinfo {volume} {D89}},\ \bibinfo {pages}
  {085024} (\bibinfo {year} {2014})},\ \Eprint {http://arxiv.org/abs/1401.6972}
  {arXiv:1401.6972 [hep-th]} \BibitemShut {NoStop}%
\bibitem [{\citenamefont {Gursoy}\ \emph {et~al.}(2014)\citenamefont {Gursoy},
  \citenamefont {Kharzeev},\ and\ \citenamefont {Rajagopal}}]{Gursoy:2014aka}%
  \BibitemOpen
  \bibfield  {author} {\bibinfo {author} {\bibfnamefont {U.}~\bibnamefont
  {Gursoy}}, \bibinfo {author} {\bibfnamefont {D.}~\bibnamefont {Kharzeev}}, \
  and\ \bibinfo {author} {\bibfnamefont {K.}~\bibnamefont {Rajagopal}},\ }\href
  {\doibase 10.1103/PhysRevC.89.054905} {\bibfield  {journal} {\bibinfo
  {journal} {Phys. Rev.}\ }\textbf {\bibinfo {volume} {C89}},\ \bibinfo {pages}
  {054905} (\bibinfo {year} {2014})},\ \Eprint {http://arxiv.org/abs/1401.3805}
  {arXiv:1401.3805 [hep-ph]} \BibitemShut {NoStop}%
\bibitem [{\citenamefont {Pu}\ \emph {et~al.}(2015)\citenamefont {Pu},
  \citenamefont {Wu},\ and\ \citenamefont {Yang}}]{Pu:2014fva}%
  \BibitemOpen
  \bibfield  {author} {\bibinfo {author} {\bibfnamefont {S.}~\bibnamefont
  {Pu}}, \bibinfo {author} {\bibfnamefont {S.-Y.}\ \bibnamefont {Wu}}, \ and\
  \bibinfo {author} {\bibfnamefont {D.-L.}\ \bibnamefont {Yang}},\ }\href
  {\doibase 10.1103/PhysRevD.91.025011} {\bibfield  {journal} {\bibinfo
  {journal} {Phys. Rev.}\ }\textbf {\bibinfo {volume} {D91}},\ \bibinfo {pages}
  {025011} (\bibinfo {year} {2015})},\ \Eprint {http://arxiv.org/abs/1407.3168}
  {arXiv:1407.3168 [hep-th]} \BibitemShut {NoStop}%
\bibitem [{\citenamefont {Chen}\ \emph {et~al.}(2016)\citenamefont {Chen},
  \citenamefont {Ishii}, \citenamefont {Pu},\ and\ \citenamefont
  {Yamamoto}}]{Chen:2016xtg}%
  \BibitemOpen
  \bibfield  {author} {\bibinfo {author} {\bibfnamefont {J.-W.}\ \bibnamefont
  {Chen}}, \bibinfo {author} {\bibfnamefont {T.}~\bibnamefont {Ishii}},
  \bibinfo {author} {\bibfnamefont {S.}~\bibnamefont {Pu}}, \ and\ \bibinfo
  {author} {\bibfnamefont {N.}~\bibnamefont {Yamamoto}},\ }\href {\doibase
  10.1103/PhysRevD.93.125023} {\bibfield  {journal} {\bibinfo  {journal} {Phys.
  Rev.}\ }\textbf {\bibinfo {volume} {D93}},\ \bibinfo {pages} {125023}
  (\bibinfo {year} {2016})},\ \Eprint {http://arxiv.org/abs/1603.03620}
  {arXiv:1603.03620 [hep-th]} \BibitemShut {NoStop}%
\bibitem [{\citenamefont {Gorbar}\ \emph {et~al.}(2016)\citenamefont {Gorbar},
  \citenamefont {Shovkovy}, \citenamefont {Vilchinskii}, \citenamefont
  {Rudenok}, \citenamefont {Boyarsky},\ and\ \citenamefont
  {Ruchayskiy}}]{Gorbar:2016qfh}%
  \BibitemOpen
  \bibfield  {author} {\bibinfo {author} {\bibfnamefont {E.~V.}\ \bibnamefont
  {Gorbar}}, \bibinfo {author} {\bibfnamefont {I.~A.}\ \bibnamefont
  {Shovkovy}}, \bibinfo {author} {\bibfnamefont {S.}~\bibnamefont
  {Vilchinskii}}, \bibinfo {author} {\bibfnamefont {I.}~\bibnamefont
  {Rudenok}}, \bibinfo {author} {\bibfnamefont {A.}~\bibnamefont {Boyarsky}}, \
  and\ \bibinfo {author} {\bibfnamefont {O.}~\bibnamefont {Ruchayskiy}},\
  }\href {\doibase 10.1103/PhysRevD.93.105028} {\bibfield  {journal} {\bibinfo
  {journal} {Phys. Rev.}\ }\textbf {\bibinfo {volume} {D93}},\ \bibinfo {pages}
  {105028} (\bibinfo {year} {2016})},\ \Eprint
  {http://arxiv.org/abs/1603.03442} {arXiv:1603.03442 [hep-th]} \BibitemShut
  {NoStop}%
\bibitem [{\citenamefont {Mace}\ \emph {et~al.}(2016)\citenamefont {Mace},
  \citenamefont {Schlichting},\ and\ \citenamefont
  {Venugopalan}}]{Mace:2016svc}%
  \BibitemOpen
  \bibfield  {author} {\bibinfo {author} {\bibfnamefont {M.}~\bibnamefont
  {Mace}}, \bibinfo {author} {\bibfnamefont {S.}~\bibnamefont {Schlichting}}, \
  and\ \bibinfo {author} {\bibfnamefont {R.}~\bibnamefont {Venugopalan}},\
  }\href {\doibase 10.1103/PhysRevD.93.074036} {\bibfield  {journal} {\bibinfo
  {journal} {Phys. Rev.}\ }\textbf {\bibinfo {volume} {D93}},\ \bibinfo {pages}
  {074036} (\bibinfo {year} {2016})},\ \Eprint
  {http://arxiv.org/abs/1601.07342} {arXiv:1601.07342 [hep-ph]} \BibitemShut
  {NoStop}%
\bibitem [{\citenamefont {Mace}\ \emph {et~al.}(2017)\citenamefont {Mace},
  \citenamefont {Mueller}, \citenamefont {Schlichting},\ and\ \citenamefont
  {Sharma}}]{Mace:2016shq}%
  \BibitemOpen
  \bibfield  {author} {\bibinfo {author} {\bibfnamefont {M.}~\bibnamefont
  {Mace}}, \bibinfo {author} {\bibfnamefont {N.}~\bibnamefont {Mueller}},
  \bibinfo {author} {\bibfnamefont {S.}~\bibnamefont {Schlichting}}, \ and\
  \bibinfo {author} {\bibfnamefont {S.}~\bibnamefont {Sharma}},\ }\href
  {\doibase 10.1103/PhysRevD.95.036023} {\bibfield  {journal} {\bibinfo
  {journal} {Phys. Rev.}\ }\textbf {\bibinfo {volume} {D95}},\ \bibinfo {pages}
  {036023} (\bibinfo {year} {2017})},\ \Eprint
  {http://arxiv.org/abs/1612.02477} {arXiv:1612.02477 [hep-lat]} \BibitemShut
  {NoStop}%
\bibitem [{\citenamefont {Berges}\ \emph {et~al.}(2017)\citenamefont {Berges},
  \citenamefont {Mace},\ and\ \citenamefont {Schlichting}}]{Berges:2017igc}%
  \BibitemOpen
  \bibfield  {author} {\bibinfo {author} {\bibfnamefont {J.}~\bibnamefont
  {Berges}}, \bibinfo {author} {\bibfnamefont {M.}~\bibnamefont {Mace}}, \ and\
  \bibinfo {author} {\bibfnamefont {S.}~\bibnamefont {Schlichting}},\
  }\href@noop {} {\  (\bibinfo {year} {2017})},\ \Eprint
  {http://arxiv.org/abs/1703.00697} {arXiv:1703.00697 [hep-th]} \BibitemShut
  {NoStop}%
\bibitem [{\citenamefont {Hirono}\ \emph {et~al.}(2014)\citenamefont {Hirono},
  \citenamefont {Hirano},\ and\ \citenamefont {Kharzeev}}]{Hirono:2014oda}%
  \BibitemOpen
  \bibfield  {author} {\bibinfo {author} {\bibfnamefont {Y.}~\bibnamefont
  {Hirono}}, \bibinfo {author} {\bibfnamefont {T.}~\bibnamefont {Hirano}}, \
  and\ \bibinfo {author} {\bibfnamefont {D.~E.}\ \bibnamefont {Kharzeev}},\
  }\href@noop {} {\  (\bibinfo {year} {2014})},\ \Eprint
  {http://arxiv.org/abs/1412.0311} {arXiv:1412.0311 [hep-ph]} \BibitemShut
  {NoStop}%
\bibitem [{\citenamefont {Jiang}\ \emph {et~al.}(2016)\citenamefont {Jiang},
  \citenamefont {Shi}, \citenamefont {Yin},\ and\ \citenamefont
  {Liao}}]{Jiang:2016wve}%
  \BibitemOpen
  \bibfield  {author} {\bibinfo {author} {\bibfnamefont {Y.}~\bibnamefont
  {Jiang}}, \bibinfo {author} {\bibfnamefont {S.}~\bibnamefont {Shi}}, \bibinfo
  {author} {\bibfnamefont {Y.}~\bibnamefont {Yin}}, \ and\ \bibinfo {author}
  {\bibfnamefont {J.}~\bibnamefont {Liao}},\ }\href@noop {} {\  (\bibinfo
  {year} {2016})},\ \Eprint {http://arxiv.org/abs/1611.04586} {arXiv:1611.04586
  [nucl-th]} \BibitemShut {NoStop}%
\bibitem [{\citenamefont {Roy}\ \emph {et~al.}(2015)\citenamefont {Roy},
  \citenamefont {Pu}, \citenamefont {Rezzolla},\ and\ \citenamefont
  {Rischke}}]{Roy:2015kma}%
  \BibitemOpen
  \bibfield  {author} {\bibinfo {author} {\bibfnamefont {V.}~\bibnamefont
  {Roy}}, \bibinfo {author} {\bibfnamefont {S.}~\bibnamefont {Pu}}, \bibinfo
  {author} {\bibfnamefont {L.}~\bibnamefont {Rezzolla}}, \ and\ \bibinfo
  {author} {\bibfnamefont {D.}~\bibnamefont {Rischke}},\ }\href {\doibase
  10.1016/j.physletb.2015.08.046} {\bibfield  {journal} {\bibinfo  {journal}
  {Phys. Lett.}\ }\textbf {\bibinfo {volume} {B750}},\ \bibinfo {pages} {45}
  (\bibinfo {year} {2015})},\ \Eprint {http://arxiv.org/abs/1506.06620}
  {arXiv:1506.06620 [nucl-th]} \BibitemShut {NoStop}%
\bibitem [{\citenamefont {Pu}\ \emph {et~al.}(2016)\citenamefont {Pu},
  \citenamefont {Roy}, \citenamefont {Rezzolla},\ and\ \citenamefont
  {Rischke}}]{Pu:2016ayh}%
  \BibitemOpen
  \bibfield  {author} {\bibinfo {author} {\bibfnamefont {S.}~\bibnamefont
  {Pu}}, \bibinfo {author} {\bibfnamefont {V.}~\bibnamefont {Roy}}, \bibinfo
  {author} {\bibfnamefont {L.}~\bibnamefont {Rezzolla}}, \ and\ \bibinfo
  {author} {\bibfnamefont {D.~H.}\ \bibnamefont {Rischke}},\ }\href {\doibase
  10.1103/PhysRevD.93.074022} {\bibfield  {journal} {\bibinfo  {journal} {Phys.
  Rev.}\ }\textbf {\bibinfo {volume} {D93}},\ \bibinfo {pages} {074022}
  (\bibinfo {year} {2016})},\ \Eprint {http://arxiv.org/abs/1602.04953}
  {arXiv:1602.04953 [nucl-th]} \BibitemShut {NoStop}%
\bibitem [{\citenamefont {Pu}\ and\ \citenamefont {Yang}(2016)}]{Pu:2016bxy}%
  \BibitemOpen
  \bibfield  {author} {\bibinfo {author} {\bibfnamefont {S.}~\bibnamefont
  {Pu}}\ and\ \bibinfo {author} {\bibfnamefont {D.-L.}\ \bibnamefont {Yang}},\
  }\href {\doibase 10.1103/PhysRevD.93.054042} {\bibfield  {journal} {\bibinfo
  {journal} {Phys. Rev.}\ }\textbf {\bibinfo {volume} {D93}},\ \bibinfo {pages}
  {054042} (\bibinfo {year} {2016})},\ \Eprint
  {http://arxiv.org/abs/1602.04954} {arXiv:1602.04954 [nucl-th]} \BibitemShut
  {NoStop}%
\bibitem [{\citenamefont {Inghirami}\ \emph {et~al.}(2016)\citenamefont
  {Inghirami}, \citenamefont {Del~Zanna}, \citenamefont {Beraudo},
  \citenamefont {Moghaddam}, \citenamefont {Becattini},\ and\ \citenamefont
  {Bleicher}}]{Inghirami:2016iru}%
  \BibitemOpen
  \bibfield  {author} {\bibinfo {author} {\bibfnamefont {G.}~\bibnamefont
  {Inghirami}}, \bibinfo {author} {\bibfnamefont {L.}~\bibnamefont
  {Del~Zanna}}, \bibinfo {author} {\bibfnamefont {A.}~\bibnamefont {Beraudo}},
  \bibinfo {author} {\bibfnamefont {M.~H.}\ \bibnamefont {Moghaddam}}, \bibinfo
  {author} {\bibfnamefont {F.}~\bibnamefont {Becattini}}, \ and\ \bibinfo
  {author} {\bibfnamefont {M.}~\bibnamefont {Bleicher}},\ }\href {\doibase
  10.1140/epjc/s10052-016-4516-8} {\bibfield  {journal} {\bibinfo  {journal}
  {Eur. Phys. J.}\ }\textbf {\bibinfo {volume} {C76}},\ \bibinfo {pages} {659}
  (\bibinfo {year} {2016})},\ \Eprint {http://arxiv.org/abs/1609.03042}
  {arXiv:1609.03042 [hep-ph]} \BibitemShut {NoStop}%
\bibitem [{\citenamefont {Sun}\ \emph {et~al.}(2016)\citenamefont {Sun},
  \citenamefont {Ko},\ and\ \citenamefont {Li}}]{Sun:2016nig}%
  \BibitemOpen
  \bibfield  {author} {\bibinfo {author} {\bibfnamefont {Y.}~\bibnamefont
  {Sun}}, \bibinfo {author} {\bibfnamefont {C.~M.}\ \bibnamefont {Ko}}, \ and\
  \bibinfo {author} {\bibfnamefont {F.}~\bibnamefont {Li}},\ }\href {\doibase
  10.1103/PhysRevC.94.045204} {\bibfield  {journal} {\bibinfo  {journal} {Phys.
  Rev.}\ }\textbf {\bibinfo {volume} {C94}},\ \bibinfo {pages} {045204}
  (\bibinfo {year} {2016})},\ \Eprint {http://arxiv.org/abs/1606.05627}
  {arXiv:1606.05627 [nucl-th]} \BibitemShut {NoStop}%
\bibitem [{\citenamefont {Huang}\ \emph {et~al.}(2017)\citenamefont {Huang},
  \citenamefont {Jiang}, \citenamefont {Shi}, \citenamefont {Liao},\ and\
  \citenamefont {Zhuang}}]{Huang:2017tsq}%
  \BibitemOpen
  \bibfield  {author} {\bibinfo {author} {\bibfnamefont {A.}~\bibnamefont
  {Huang}}, \bibinfo {author} {\bibfnamefont {Y.}~\bibnamefont {Jiang}},
  \bibinfo {author} {\bibfnamefont {S.}~\bibnamefont {Shi}}, \bibinfo {author}
  {\bibfnamefont {J.}~\bibnamefont {Liao}}, \ and\ \bibinfo {author}
  {\bibfnamefont {P.}~\bibnamefont {Zhuang}},\ }\href@noop {} {\  (\bibinfo
  {year} {2017})},\ \Eprint {http://arxiv.org/abs/1703.08856} {arXiv:1703.08856
  [hep-ph]} \BibitemShut {NoStop}%
\bibitem [{\citenamefont {Berry}(1984)}]{Berry1984}%
  \BibitemOpen
  \bibfield  {author} {\bibinfo {author} {\bibfnamefont {M.~V.}\ \bibnamefont
  {Berry}},\ }\href {\doibase 10.1098/rspa.1984.0023} {\bibfield  {journal}
  {\bibinfo  {journal} {Proc. Roy. Soc. Lond.}\ }\textbf {\bibinfo {volume}
  {A392}},\ \bibinfo {pages} {45} (\bibinfo {year} {1984})}\BibitemShut
  {NoStop}%
\bibitem [{\citenamefont {Xiao}\ \emph {et~al.}(2005)\citenamefont {Xiao},
  \citenamefont {Shi},\ and\ \citenamefont {Niu}}]{Xiao:2005eif}%
  \BibitemOpen
  \bibfield  {author} {\bibinfo {author} {\bibfnamefont {D.}~\bibnamefont
  {Xiao}}, \bibinfo {author} {\bibfnamefont {J.}~\bibnamefont {Shi}}, \ and\
  \bibinfo {author} {\bibfnamefont {Q.}~\bibnamefont {Niu}},\ }\href {\doibase
  10.1103/PhysRevLett.95.137204} {\bibfield  {journal} {\bibinfo  {journal}
  {Phys. Rev. Lett.}\ }\textbf {\bibinfo {volume} {95}},\ \bibinfo {pages}
  {137204} (\bibinfo {year} {2005})}\BibitemShut {NoStop}%
\bibitem [{\citenamefont {Duval}\ \emph {et~al.}(2006)\citenamefont {Duval},
  \citenamefont {Horvath}, \citenamefont {Horvathy}, \citenamefont {Martina},\
  and\ \citenamefont {Stichel}}]{Duval2006}%
  \BibitemOpen
  \bibfield  {author} {\bibinfo {author} {\bibfnamefont {C.}~\bibnamefont
  {Duval}}, \bibinfo {author} {\bibfnamefont {Z.}~\bibnamefont {Horvath}},
  \bibinfo {author} {\bibfnamefont {P.}~\bibnamefont {Horvathy}}, \bibinfo
  {author} {\bibfnamefont {L.}~\bibnamefont {Martina}}, \ and\ \bibinfo
  {author} {\bibfnamefont {P.}~\bibnamefont {Stichel}},\ }\href {\doibase
  10.1142/S0217984906010573} {\bibfield  {journal} {\bibinfo  {journal}
  {Mod.Phys.Lett.}\ }\textbf {\bibinfo {volume} {B20}},\ \bibinfo {pages} {373}
  (\bibinfo {year} {2006})},\ \Eprint {http://arxiv.org/abs/cond-mat/0506051}
  {arXiv:cond-mat/0506051 [cond-mat]} \BibitemShut {NoStop}%
\bibitem [{\citenamefont {Xiao}\ \emph {et~al.}(2010)\citenamefont {Xiao},
  \citenamefont {Chang},\ and\ \citenamefont {Niu}}]{Xiao2010}%
  \BibitemOpen
  \bibfield  {author} {\bibinfo {author} {\bibfnamefont {D.}~\bibnamefont
  {Xiao}}, \bibinfo {author} {\bibfnamefont {M.-C.}\ \bibnamefont {Chang}}, \
  and\ \bibinfo {author} {\bibfnamefont {Q.}~\bibnamefont {Niu}},\ }\href
  {\doibase 10.1103/RevModPhys.82.1959} {\bibfield  {journal} {\bibinfo
  {journal} {Rev. Mod. Phys.}\ }\textbf {\bibinfo {volume} {82}},\ \bibinfo
  {pages} {1959} (\bibinfo {year} {2010})},\ \Eprint
  {http://arxiv.org/abs/0907.2021} {arXiv:0907.2021 [cond-mat.mes-hall]}
  \BibitemShut {NoStop}%
\bibitem [{\citenamefont {Chen}\ \emph
  {et~al.}(2014{\natexlab{a}})\citenamefont {Chen}, \citenamefont {Pang},
  \citenamefont {Pu},\ and\ \citenamefont {Wang}}]{Chen:2013iga}%
  \BibitemOpen
  \bibfield  {author} {\bibinfo {author} {\bibfnamefont {J.-W.}\ \bibnamefont
  {Chen}}, \bibinfo {author} {\bibfnamefont {J.-y.}\ \bibnamefont {Pang}},
  \bibinfo {author} {\bibfnamefont {S.}~\bibnamefont {Pu}}, \ and\ \bibinfo
  {author} {\bibfnamefont {Q.}~\bibnamefont {Wang}},\ }\href {\doibase
  10.1103/PhysRevD.89.094003} {\bibfield  {journal} {\bibinfo  {journal} {Phys.
  Rev.}\ }\textbf {\bibinfo {volume} {D89}},\ \bibinfo {pages} {094003}
  (\bibinfo {year} {2014}{\natexlab{a}})},\ \Eprint
  {http://arxiv.org/abs/1312.2032} {arXiv:1312.2032 [hep-th]} \BibitemShut
  {NoStop}%
\bibitem [{\citenamefont {Chen}\ \emph
  {et~al.}(2014{\natexlab{b}})\citenamefont {Chen}, \citenamefont {Son},
  \citenamefont {Stephanov}, \citenamefont {Yee},\ and\ \citenamefont
  {Yin}}]{Chen:2014cla}%
  \BibitemOpen
  \bibfield  {author} {\bibinfo {author} {\bibfnamefont {J.-Y.}\ \bibnamefont
  {Chen}}, \bibinfo {author} {\bibfnamefont {D.~T.}\ \bibnamefont {Son}},
  \bibinfo {author} {\bibfnamefont {M.~A.}\ \bibnamefont {Stephanov}}, \bibinfo
  {author} {\bibfnamefont {H.-U.}\ \bibnamefont {Yee}}, \ and\ \bibinfo
  {author} {\bibfnamefont {Y.}~\bibnamefont {Yin}},\ }\href {\doibase
  10.1103/PhysRevLett.113.182302} {\bibfield  {journal} {\bibinfo  {journal}
  {Phys. Rev. Lett.}\ }\textbf {\bibinfo {volume} {113}},\ \bibinfo {pages}
  {182302} (\bibinfo {year} {2014}{\natexlab{b}})},\ \Eprint
  {http://arxiv.org/abs/1404.5963} {arXiv:1404.5963 [hep-th]} \BibitemShut
  {NoStop}%
\bibitem [{\citenamefont {Hidaka}\ \emph {et~al.}(2017)\citenamefont {Hidaka},
  \citenamefont {Pu},\ and\ \citenamefont {Yang}}]{Hidaka:2016yjf}%
  \BibitemOpen
  \bibfield  {author} {\bibinfo {author} {\bibfnamefont {Y.}~\bibnamefont
  {Hidaka}}, \bibinfo {author} {\bibfnamefont {S.}~\bibnamefont {Pu}}, \ and\
  \bibinfo {author} {\bibfnamefont {D.-L.}\ \bibnamefont {Yang}},\ }\href
  {\doibase 10.1103/PhysRevD.95.091901} {\bibfield  {journal} {\bibinfo
  {journal} {Phys. Rev.}\ }\textbf {\bibinfo {volume} {D95}},\ \bibinfo {pages}
  {091901} (\bibinfo {year} {2017})},\ \Eprint
  {http://arxiv.org/abs/1612.04630} {arXiv:1612.04630 [hep-th]} \BibitemShut
  {NoStop}%
\bibitem [{\citenamefont {Baym}(1984)}]{Baym:1984guh}%
  \BibitemOpen
  \bibfield  {author} {\bibinfo {author} {\bibfnamefont {G.}~\bibnamefont
  {Baym}},\ }\href {\doibase http://dx.doi.org/10.1016/0370-2693(84)91863-X}
  {\bibfield  {journal} {\bibinfo  {journal} {Physics Letters B}\ }\textbf
  {\bibinfo {volume} {138}},\ \bibinfo {pages} {18 } (\bibinfo {year}
  {1984})}\BibitemShut {NoStop}%
\bibitem [{\citenamefont {Mueller}(2000)}]{Mueller:1999pi}%
  \BibitemOpen
  \bibfield  {author} {\bibinfo {author} {\bibfnamefont {A.~H.}\ \bibnamefont
  {Mueller}},\ }\href {\doibase 10.1016/S0370-2693(00)00084-8} {\bibfield
  {journal} {\bibinfo  {journal} {Phys. Lett.}\ }\textbf {\bibinfo {volume}
  {B475}},\ \bibinfo {pages} {220} (\bibinfo {year} {2000})},\ \Eprint
  {http://arxiv.org/abs/hep-ph/9909388} {arXiv:hep-ph/9909388 [hep-ph]}
  \BibitemShut {NoStop}%
\bibitem [{\citenamefont {Baier}\ \emph {et~al.}(2001)\citenamefont {Baier},
  \citenamefont {Mueller}, \citenamefont {Schiff},\ and\ \citenamefont
  {Son}}]{Baier:2000sb}%
  \BibitemOpen
  \bibfield  {author} {\bibinfo {author} {\bibfnamefont {R.}~\bibnamefont
  {Baier}}, \bibinfo {author} {\bibfnamefont {A.~H.}\ \bibnamefont {Mueller}},
  \bibinfo {author} {\bibfnamefont {D.}~\bibnamefont {Schiff}}, \ and\ \bibinfo
  {author} {\bibfnamefont {D.~T.}\ \bibnamefont {Son}},\ }\href {\doibase
  10.1016/S0370-2693(01)00191-5} {\bibfield  {journal} {\bibinfo  {journal}
  {Phys. Lett.}\ }\textbf {\bibinfo {volume} {B502}},\ \bibinfo {pages} {51}
  (\bibinfo {year} {2001})},\ \Eprint {http://arxiv.org/abs/hep-ph/0009237}
  {arXiv:hep-ph/0009237 [hep-ph]} \BibitemShut {NoStop}%
\bibitem [{\citenamefont {Fukushima}(2017)}]{Fukushima:2016xgg}%
  \BibitemOpen
  \bibfield  {author} {\bibinfo {author} {\bibfnamefont {K.}~\bibnamefont
  {Fukushima}},\ }\href {\doibase 10.1088/1361-6633/80/2/022301} {\bibfield
  {journal} {\bibinfo  {journal} {Rept. Prog. Phys.}\ }\textbf {\bibinfo
  {volume} {80}},\ \bibinfo {pages} {022301} (\bibinfo {year} {2017})},\
  \Eprint {http://arxiv.org/abs/1603.02340} {arXiv:1603.02340 [nucl-th]}
  \BibitemShut {NoStop}%
\bibitem [{\citenamefont {Blaizot}(2017)}]{Blaizot:2016qgz}%
  \BibitemOpen
  \bibfield  {author} {\bibinfo {author} {\bibfnamefont {J.-P.}\ \bibnamefont
  {Blaizot}},\ }\href {\doibase 10.1088/1361-6633/aa5435} {\bibfield  {journal}
  {\bibinfo  {journal} {Rept. Prog. Phys.}\ }\textbf {\bibinfo {volume} {80}},\
  \bibinfo {pages} {032301} (\bibinfo {year} {2017})},\ \Eprint
  {http://arxiv.org/abs/1607.04448} {arXiv:1607.04448 [hep-ph]} \BibitemShut
  {NoStop}%
\bibitem [{Note1()}]{Note1}%
  \BibitemOpen
  \bibinfo {note} {We use a word ``quantum'' in a sense of discussions in
  Refs.~\cite {Epelbaum:2014mfa,Fukushima:2017odi}.}\BibitemShut {Stop}%
\bibitem [{\citenamefont {Kharzeev}\ \emph {et~al.}(2002)\citenamefont
  {Kharzeev}, \citenamefont {Krasnitz},\ and\ \citenamefont
  {Venugopalan}}]{Kharzeev:2001ev}%
  \BibitemOpen
  \bibfield  {author} {\bibinfo {author} {\bibfnamefont {D.}~\bibnamefont
  {Kharzeev}}, \bibinfo {author} {\bibfnamefont {A.}~\bibnamefont {Krasnitz}},
  \ and\ \bibinfo {author} {\bibfnamefont {R.}~\bibnamefont {Venugopalan}},\
  }\href {\doibase 10.1016/S0370-2693(02)02630-8} {\bibfield  {journal}
  {\bibinfo  {journal} {Phys. Lett.}\ }\textbf {\bibinfo {volume} {B545}},\
  \bibinfo {pages} {298} (\bibinfo {year} {2002})},\ \Eprint
  {http://arxiv.org/abs/hep-ph/0109253} {arXiv:hep-ph/0109253 [hep-ph]}
  \BibitemShut {NoStop}%
\bibitem [{\citenamefont {Chen}\ \emph {et~al.}(2015)\citenamefont {Chen},
  \citenamefont {Son},\ and\ \citenamefont {Stephanov}}]{Chen:2015jop}%
  \BibitemOpen
  \bibfield  {author} {\bibinfo {author} {\bibfnamefont {J.-Y.}\ \bibnamefont
  {Chen}}, \bibinfo {author} {\bibfnamefont {D.~T.}\ \bibnamefont {Son}}, \
  and\ \bibinfo {author} {\bibfnamefont {M.~A.}\ \bibnamefont {Stephanov}},\
  }\href {\doibase 10.1103/PhysRevLett.115.021601} {\bibfield  {journal}
  {\bibinfo  {journal} {Phys. Rev. Lett.}\ }\textbf {\bibinfo {volume} {115}},\
  \bibinfo {pages} {021601} (\bibinfo {year} {2015})},\ \Eprint
  {http://arxiv.org/abs/1502.06966} {arXiv:1502.06966 [hep-th]} \BibitemShut
  {NoStop}%
\bibitem [{\citenamefont {Romatschke}\ and\ \citenamefont
  {Venugopalan}(2006)}]{Romatschke:2006nk}%
  \BibitemOpen
  \bibfield  {author} {\bibinfo {author} {\bibfnamefont {P.}~\bibnamefont
  {Romatschke}}\ and\ \bibinfo {author} {\bibfnamefont {R.}~\bibnamefont
  {Venugopalan}},\ }\href {\doibase 10.1103/PhysRevD.74.045011} {\bibfield
  {journal} {\bibinfo  {journal} {Phys. Rev.}\ }\textbf {\bibinfo {volume}
  {D74}},\ \bibinfo {pages} {045011} (\bibinfo {year} {2006})},\ \Eprint
  {http://arxiv.org/abs/hep-ph/0605045} {arXiv:hep-ph/0605045 [hep-ph]}
  \BibitemShut {NoStop}%
\bibitem [{\citenamefont {Gelis}\ \emph {et~al.}(2005)\citenamefont {Gelis},
  \citenamefont {Kajantie},\ and\ \citenamefont {Lappi}}]{Gelis:2004jp}%
  \BibitemOpen
  \bibfield  {author} {\bibinfo {author} {\bibfnamefont {F.}~\bibnamefont
  {Gelis}}, \bibinfo {author} {\bibfnamefont {K.}~\bibnamefont {Kajantie}}, \
  and\ \bibinfo {author} {\bibfnamefont {T.}~\bibnamefont {Lappi}},\ }\href
  {\doibase 10.1103/PhysRevC.71.024904} {\bibfield  {journal} {\bibinfo
  {journal} {Phys. Rev.}\ }\textbf {\bibinfo {volume} {C71}},\ \bibinfo {pages}
  {024904} (\bibinfo {year} {2005})},\ \Eprint
  {http://arxiv.org/abs/hep-ph/0409058} {arXiv:hep-ph/0409058 [hep-ph]}
  \BibitemShut {NoStop}%
\bibitem [{\citenamefont {Gelis}\ and\ \citenamefont
  {Tanji}(2016)}]{Gelis:2015eua}%
  \BibitemOpen
  \bibfield  {author} {\bibinfo {author} {\bibfnamefont {F.}~\bibnamefont
  {Gelis}}\ and\ \bibinfo {author} {\bibfnamefont {N.}~\bibnamefont {Tanji}},\
  }\href {\doibase 10.1007/JHEP02(2016)126} {\bibfield  {journal} {\bibinfo
  {journal} {JHEP}\ }\textbf {\bibinfo {volume} {02}},\ \bibinfo {pages} {126}
  (\bibinfo {year} {2016})},\ \Eprint {http://arxiv.org/abs/1506.03327}
  {arXiv:1506.03327 [hep-ph]} \BibitemShut {NoStop}%
\bibitem [{\citenamefont {Epelbaum}\ \emph {et~al.}(2014)\citenamefont
  {Epelbaum}, \citenamefont {Gelis}, \citenamefont {Tanji},\ and\ \citenamefont
  {Wu}}]{Epelbaum:2014mfa}%
  \BibitemOpen
  \bibfield  {author} {\bibinfo {author} {\bibfnamefont {T.}~\bibnamefont
  {Epelbaum}}, \bibinfo {author} {\bibfnamefont {F.}~\bibnamefont {Gelis}},
  \bibinfo {author} {\bibfnamefont {N.}~\bibnamefont {Tanji}}, \ and\ \bibinfo
  {author} {\bibfnamefont {B.}~\bibnamefont {Wu}},\ }\href {\doibase
  10.1103/PhysRevD.90.125032} {\bibfield  {journal} {\bibinfo  {journal} {Phys.
  Rev.}\ }\textbf {\bibinfo {volume} {D90}},\ \bibinfo {pages} {125032}
  (\bibinfo {year} {2014})},\ \Eprint {http://arxiv.org/abs/1409.0701}
  {arXiv:1409.0701 [hep-ph]} \BibitemShut {NoStop}%
\bibitem [{\citenamefont {Fukushima}\ \emph {et~al.}(2017)\citenamefont
  {Fukushima}, \citenamefont {Murase},\ and\ \citenamefont
  {Pu}}]{Fukushima:2017odi}%
  \BibitemOpen
  \bibfield  {author} {\bibinfo {author} {\bibfnamefont {K.}~\bibnamefont
  {Fukushima}}, \bibinfo {author} {\bibfnamefont {K.}~\bibnamefont {Murase}}, \
  and\ \bibinfo {author} {\bibfnamefont {S.}~\bibnamefont {Pu}},\ }\href@noop
  {} {\  (\bibinfo {year} {2017})},\ \Eprint {http://arxiv.org/abs/1703.09492}
  {arXiv:1703.09492 [hep-ph]} \BibitemShut {NoStop}%
\end{thebibliography}%
\bibliographystyle{apsrev4-1}

\end{document}